\documentclass[10pt,aps,prd,fleqn,superscriptaddress,notitlepage,nofootinbib,preprintnumbers,nobalancelastpage]{revtex4-1}
\pdfoutput=1
\usepackage{amsmath,amssymb,graphicx,xspace}
\usepackage[caption=false]{subfig}
\usepackage{hyperref,xcolor,listings}
\newcommand{\sherpa}{\textsc{Sherpa}\xspace}

\newcommand{\bjet}{\ensuremath{b\text{-jet}}}
\newcommand{\bbjet}{\ensuremath{b\bar{b}\text{-jet}}}

\newcommand{\ttbb}{\ensuremath{\ttbar\bbbar}\xspace}
\newcommand{\ttj}{\ensuremath{\ttbar\text{+jets}}\xspace}
\newcommand{\bquark}{\ensuremath{b\text{-quark}}}

\newcommand{\bbbar}{\ensuremath{b\bar{b}}}
\newcommand{\tth}{\ensuremath{\ttbar H}}
\newcommand{\tthbb}{\ensuremath{t\bar{t}[H\to b\bar{b}]}}
\newcommand{\ttbar}{\ensuremath{t\bar{t}}\xspace}

\newcommand{\oneb}{\ensuremath{\ttbar + \ge 1 \bjet}\xspace}
\newcommand{\twob}{\ensuremath{\ttbar + \ge 2} \bjet s\xspace}
\newcommand{\ttc}{\ensuremath{\ttbar + \ge 1} charm jet\xspace}
\newcommand{\pt}{$p_\perp$\,}

\hypersetup{
  pdfauthor={Lars Ferencz,Stefan Hoeche,Judith Katzy,Frank Siegert},
  pdftitle={Simulating ttbb at NLO precision in a variable flavor number scheme}
}

\begin{document}
\preprint{DESY-24-022, FERMILAB-PUB-24-024-T, MCNET-24-02}
\title{\texorpdfstring{\boldmath$t\bar{t}b\bar{b}$}{ttbb} at NLO precision in a variable flavor number scheme}
\author{Lars Ferencz}
\affiliation{Deutsches Elektronen-Synchrotron DESY, 22603 Hamburg, Germany}
\author{Stefan~H{\"o}che}
\affiliation{Fermi National Accelerator Laboratory, Batavia, IL, 60510, USA}
\author{Judith Katzy}
\affiliation{Deutsches Elektronen-Synchrotron DESY, 22603 Hamburg, Germany}
\author{Frank Siegert}
\affiliation{Institute of Nuclear and Particle Physics, Technische Universit{\"a}t Dresden, 01062 Dresden, Germany}

\begin{abstract}
Top-quark pair production in association with two $b$-jets is computed at next-to-leading order
QCD precision, including effects of the $b$-quark mass, and matched to a $t\bar{t}$+jets simulation
in a variable flavor number scheme. The Monte Carlo realization of this method, called fusing,
consistently embeds the four-flavor calculation in a particle-level event generator.
As a first phenomenological application, we present observables relevant to the data-driven
estimation of irreducible backgrounds to $t\bar{t}H$-production.
\end{abstract}

\maketitle
\section{Introduction}
The production of a pair of top quarks in association with a pair of bottom quarks (\ttbb)
is notoriously difficult to model at high-energy hadron colliders. QCD effects induced by
a large top-quark mass, a comparably small bottom-quark mass and potentially large transverse
energies of additional light jets must all be described simultaneously. Understanding
the dynamics of the \ttbb\ final state is necessary because it forms a background to
most measurements and searches with multiple \bjet s at the Large Hadron Collider (LHC).
Standard Model Higgs bosons predominantly decay into $b$-quarks, making \ttbb\ 
particularly important for a direct measurement of the Higgs-top Yukawa coupling
through Higgs production with one ($tH$) or two associated top-quarks (\tth)
\cite{ATLAS:2021qou, ATLAS:2023cbt, ATLAS:2017fak,CMS:2018hnq,CMS:2018uxb}.
\ttbb\ is also relevant for di-Higgs searches \cite{ATLAS:2023qzf, CMS:2022cpr}
and measurements of four top production \cite{ATLAS:2023ajo,CMS:2023ftu}.
Similar event topologies would be observed in searches such as the one
for heavy Higgs bosons decaying into \ttbar \cite{ATLAS:2021upq}.
These analyses also suffer from background of $\ttbar+light$ and $\ttbar+charm$-jets,
as the detectors have limited capabilities to identify \bjet s, and 
the mis-identification rate increases with increasing \bjet\ efficiency. 
A simulation which is inclusive in terms of jet multiplicity and parton flavor
is therefore required to increase statistics in the search for rare processes.

The most common approach to simulating top-quark plus jets events at the LHC
is in terms of five-flavor scheme calculations matched to parton showers~\cite{
  Frixione:2002ik,Nason:2004rx,Frixione:2007vw,Alioli:2010xd,Hoeche:2010pf,
  Hoeche:2011fd,Alwall:2014hca}. These are typically performed
for a different number of jets at fixed order, and then combined using the
multi-jet merging techniques developed in~\cite{Lavesson:2008ah,Gehrmann:2012yg,
  Hoeche:2012yf,Frederix:2012ps,Lonnblad:2012ix,Platzer:2012bs}.
Individually, each of the inputs to the merging can only reliably describe
final-state configurations which do not have large scale hierarchies between
the jets at fixed-order level. Scale hierarchies generally induce logarithms
of Sudakov type, which must be resummed to all orders and are 
properly taken into account by the multi-jet merging procedure. 

Existing merging techniques are inclusive in the number of jet flavors.
Similar to their fixed-order inputs, they account for both the evolution
of the strong coupling and the running of the PDF with five active parton flavors.
This has the disadvantage that bottom quark masses are neglected,
and therefore the subtle effects of a reduced phase space and inhibited
QCD radiation in the collinear region might not be modeled correctly 
for \bjet s in the \ttbb final state. An alternative method is to compute 
\ttbb production in the four-flavor scheme~\cite{Bredenstein:2008zb,
  Bredenstein:2009aj,Bredenstein:2010rs,Buccioni:2019plc,Bevilacqua:2022twl}.
Some of the existing calculations have been matched to parton showers~\cite{
  Cascioli:2013era,Jezo:2018yaf}. The resulting predictions are inclusive
in the \ttbb phase space and cover in particular the low transverse momentum
region of all \bjet s. However, they do not model the potential production of
an additional \bjet\ through initial-state gluon splitting due to the vanishing
\bquark\ PDF. In addition, they only include Sudakov suppression to first order
in the strong coupling and cannot account for resummed higher-order effects
that set in when the hierarchy between different hard scales in the process
is large.

The four- and five-flavor scheme calculations described above can be matched
in a variable flavor number scheme such as ACOT~\cite{Aivazis:1993kh,
  Aivazis:1993pi,Tung:2001mv,Kniehl:2011bk}, TR~\cite{Thorne:1997ga,Thorne:2006qt},
or FONLL~\cite{Cacciari:1998it,Forte:2010ta}. This allows to account for both the
effect of finite parton masses at small scales, and the QCD evolution at large scales.
The equivalent of a variable flavor number scheme in a multi-jet merging context
was introduced in~\cite{Hoche:2019ncc} and named the ``fusing'' method.
For color-singlet processes, it correctly reproduces all matching coefficients 
that are relevant to the logarithmic precision of the parton shower and the precision
of the fixed-order input to the matching. In this manuscript, we present the first 
application of the fusing method to the \ttbb process, i.e.\ a process with
four colored objects in the final state. In particular, we study 
the impact on observables relevant to the measurement of \tth\  production.
The parton-shower algorithm used in our simulations
is accurate to leading color only, therefore we neither provide a complete
treatment of soft-gluon effects, nor the corresponding heavy-flavor matching.
We note, however, that sub-leading color corrections as well as sub-leading
logarithmic corrections could be included in our calculation as more precise
parton showers become available in the future.

Measurements of  the inclusive and differential cross sections
of \ttbb at a center-of-mass energy of 13\,TeV at the LHC have been performed
with partial luminosity and rather large uncertainties~\cite{CMS:2020grm,
  CMS:2019eih,ATLAS:2018fwl} and, very recently, with the full Run2 data
reaching a precision of the order of 10\%~\cite{CMS:2023xjh},
which allows to test the various theoretical predictions.  While analyses
based on the partial data set could not discriminate between different approaches
to simulating \ttbb, the more precise results will allow for stringent tests
in the future.

The outline of this manuscript is as follows:
Section~\ref{sec:merging_and_fusing} contains a basic review of the
matching, merging and fusing technology used in our approach, with a focus on
theoretical consistency of the combined event sample. In Sec.~\ref{sec:validation}
we present parton-level results obtained from the fusing procedure with stable
top quarks and discuss the implications for the QCD radiation pattern not associated
with the top decay. Finally, in Sec.~\ref{sec:results}, we present predictions
for the complete simulation, including top decays and a realistic event analysis
as well as comparisons to experimental measurements. We summarize our findings
and present an outlook in Sec.~\ref{sec:conclusions}.

\section{NLO multi-jet merging and fusing}
\label{sec:merging_and_fusing}
The incorporation of heavy-quark mass effects in the evolution of QCD partons
can be achieved through various well-established approaches~\cite{FebresCordero:2015fgt}.
General-mass variable flavor number schemes (VFNS) like
ACOT~\cite{Aivazis:1993kh,Aivazis:1993pi,Tung:2001mv,Kniehl:2011bk},
TR~\cite{Thorne:1997ga,Thorne:2006qt}, and FONLL~\cite{Cacciari:1998it,Forte:2010ta}
can account for mass effects dynamically, based on the active flavor thresholds.
They rely on the matching of heavy to light flavor matrix elements~\cite{Buza:1996wv},
and allow to consistently include calculations with massive heavy flavors
in flavor-inclusive predictions through dedicated matching coefficients.
The relation between these matching coefficients and a parton-shower algorithm
was discussed in detail in Ref.~\cite{Hoche:2019ncc}. We review only the basic
algorithm here, in order to establish the notation for the remainder of the manuscript.
Note that the \bquark s from top decays are not included for the discussion in this section,
and that we apply a narrow-width approximation in our calculations to factor out the top decays.

\subsection{Multi-jet merging at fixed number of flavors}
\label{sec:merging}
\begin{figure}
    \includegraphics[scale=0.45]{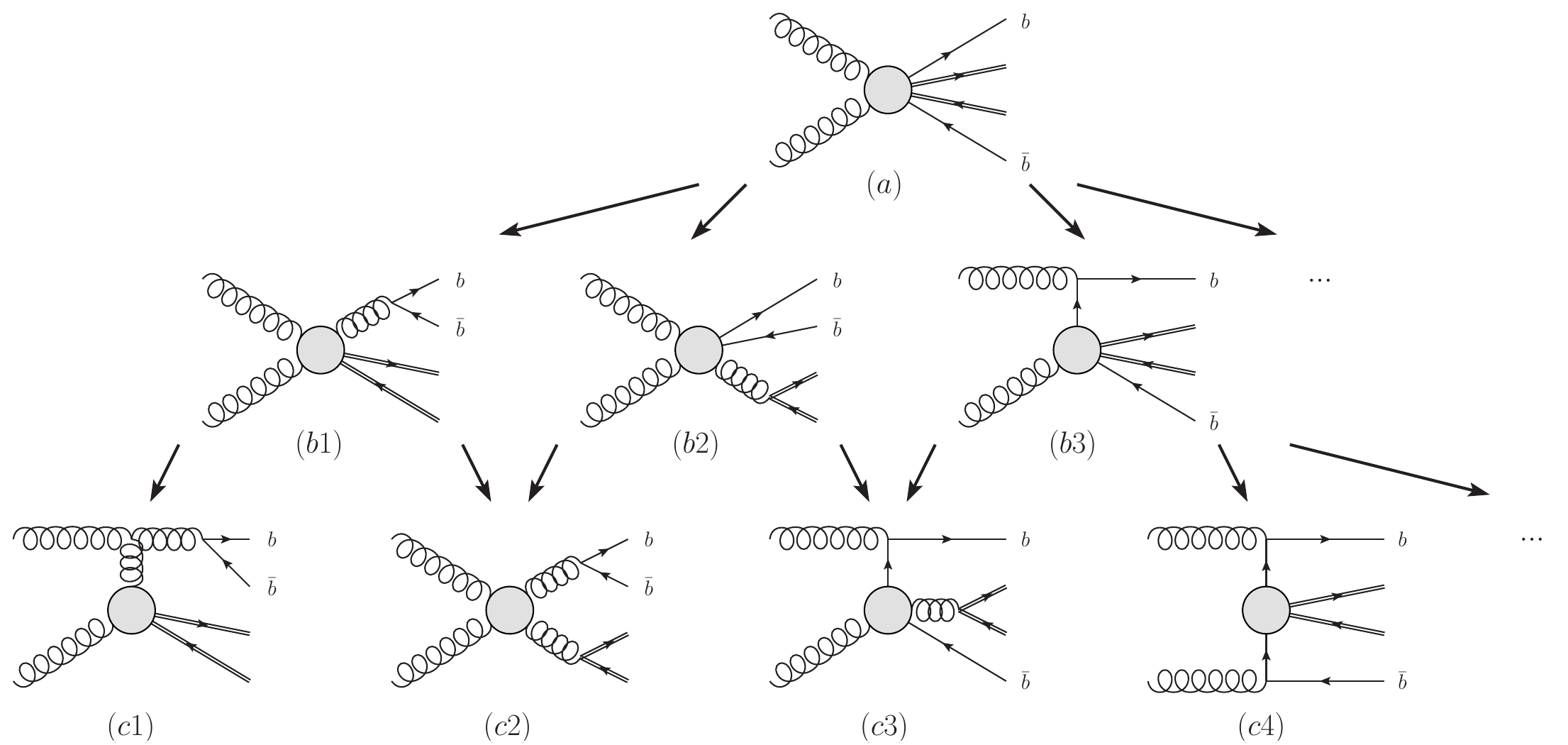}
    \caption{Examples of clustering paths and parton-shower
    histories in \ttbb\ production through gluon-gluon fusion.
    Top quarks are shown as double lines. Figure~(a) is the starting configuration,
    Figs.~(b1)-(b3) are representative configurations after the first recombination,
    and Figs.~(c1)-(c4) are representative configurations after the second recombination.
    \label{fig:clustering_examples}}
\end{figure}
Multi-jet merging has been developed as a systematic way to improve the 
accuracy of parton-shower simulations through the addition of higher-multiplicity
fixed-order calculations with well separated partons~\cite{Buckley:2011ms,Campbell:2022qmc}.
The method preserves the logarithmic accuracy of the parton shower, while predicting
configurations with hard, well-separated jets including all quantum-mechanical
correlations at the leading or next-to-leading order in QCD.
In the \sherpa event generator, these merging algorithms are implemented with
LO or NLO multi-jet matrix-elements, using either the MEPS@LO\cite{Hoeche:2009rj}
or MEPS@NLO\cite{Hoeche:2012yf,Gehrmann:2012yg} method.
The algorithm consists of various components:
\begin{enumerate}
\item {\bf Phase-space slicing}: The phase space accessible to
  multi-parton fixed-order calculations is restricted to the infrared-finite
  region, in order to make the matrix elements Monte-Carlo integrable.
  The respective cuts are performed using a {\it jet criterion}, which is
  typically based on a distance measure similar to those used in the
  longitudinally invariant $k_T$ algorithm~\cite{Catani:1991hj}.
\item\label{step:ps_history}
  {\bf Parton shower histories}: Multi-parton matrix elements are
  re-interpreted in terms of (binary) trees that represent emission
  sequences associated with a specific parton-shower event that has
  developed from a {\it core process}. These trees are called
  {\it parton-shower histories}. For each event, one history
  is selected according to the differential cross section of 
  the core process and all subsequent emissions~\cite{Andre:1997vh}.
\item {\bf \boldmath$\alpha_s$ reweighting}: The strong coupling
  is evaluated independently at each vertex of the parton-shower
  history, and includes higher-order corrections to soft gluon
  production~\cite{Amati:1980ch,Catani:1990rr}. The scale for
  residual strong coupling factors is chosen based on the 
  dynamics of the core process.
\item {\bf Sudakov weights and jet veto}: No-emission probabilities
  are computed for each intermediate parton in the parton-shower history.
  At the same time, emissions that would lead to a final state
  accessible to higher multiplicity tree-level matrix elements
  are vetoed. Typically, both steps are performed at the same time
  using pseudo-showers~\cite{Lonnblad:2001iq}.
\end{enumerate}
In the case of next-to-leading order accurate input distributions,
the algorithm is modified to account for the fact that the fixed-order
result includes the Sudakov suppression to first order in the strong
coupling~\cite{Hoeche:2012yf,Gehrmann:2012yg}.

Figure~\ref{fig:clustering_examples} shows some representative
clustering paths and parton-shower histories for the heavy-quark
associated processes \ttbb\ through gluon-gluon fusion.
The configuration in (c1) corresponds to the leading (most probable)
history in a typical LHC scenario, while the configurations
in (c2) and (c3) are strongly sub-leading. The core processes are
$gg\to t\bar{t}$~(c1) $gg\to gg$~(c2), $g\bar{b}\to g\bar{b}$~(c3)
and $b\bar{b}\to t\bar{t}$~(c4). The configurations in (c2)
and (c3) correspond to the rare situation where the parton-shower
probability for an s-channel gluon decay into a $t\bar{t}$ pair
is large, and therefore the core process differs from the naively
expected \ttbar\ process, a situation that can arise at very high
collider energies~\cite{Schalicke:2005nv}. In the LHC case,
this would require a highly boosted $t\bar{t}$ pair recoiling
against a TeV-scale \bbjet{}, such that the transverse momentum
exceeds the $t\bar{t}$ invariant mass. Events of this type, while properly
accounted for in the merging algorithm, are only relevant in extreme
regions of the phase space.

\subsection{The Fusing Algorithm}
\label{sec:fusing}
The fusing approach to heavy particle production is an intuitive
application of heavy-flavor matching, based on the ideas developed
in~\cite{Forte:2010ta,Forte:2015hba,Forte:2016sja,Forte:2018ovl}.
Massive \ttbb\ calculations are treated in the same fashion
as any other higher multiplicity input to the MEPS@NLO merging,
but no phase-space restriction is applied if no light jets are present
at Born level. The fusing technique can therefore be interpreted as
a special case of multi-jet merging, with a dedicated jet criterion
applied to heavy-flavor events. It can be shown~\cite{Hoche:2019ncc}
that this method reproduces all of the leading logarithmic matching
coefficients computed in~\cite{Buza:1996wv} for color-singlet processes.
Dedicated counterterms are needed to recover some sub-leading logarithmic
coefficients related to $\alpha_s$ renormalization. Sub-leading coefficients
arising from higher-order corrections to the DGLAP kernels are currently
not recovered, as the corresponding QCD evolution kernels are not
implemented by the parton shower. However, progress in parton-shower
development makes it likely that they will become available in the
near future~\cite{Li:2016yez,Hoche:2017hno,Dulat:2018vuy,
  Gellersen:2021eci,Campbell:2021svd,FerrarioRavasio:2023kyg}.
We will not repeat the theoretical derivation of the fusing algorithm,
but focus only on its algorithmic implementation.
This is achieved as follows:
\begin{enumerate}
\item Start with a merged simulation of the inclusive reaction
  (\ttj) and a calculation of heavy quark associated
  production (\ttbb).
\item\label{alg:fusing_2}
  Process the \ttbb events as if they were part of the inclusive,
  multi-jet merged computation, i.e.\ apply the clustering procedure, 
  the $\alpha_s$ reweighting and the Sudakov reweighting.
  Renormalization and factorization scales for the core process
  are calculated using a custom scale definition that reflects the dynamics
  of the core reaction\footnote{The core process is defined by the
  parton-shower history. In the examples of Fig.~\ref{fig:clustering_examples}
  we have the core processes $gg\to t\bar{t}$, $gg\to gg$, $g\bar{b}\to g\bar{b}$,
  and $b\bar{b}\to t\bar{t}$.} Adjust the renormalization of $\alpha_s$.
  This part of the fused result is called the {\it direct} component.
\item\label{alg:fusing_3}
  Remove all final-state configurations from the five-flavor scheme
  merged simulation of \ttj\ that have a parton-shower history
  which can also be generated at matrix-element level in the reweighted
  \ttbb\ computation.
  The remainder of the five-flavor scheme result may still contribute
  configurations with final-state bottom quarks. This part of the
  fused result is called the {\it fragmentation} component.
\item Add the modified event samples to obtain the overall prediction.
\end{enumerate}
Whether or not a configuration produced in the inclusive \ttj\
sample is of fragmentation type can be determined only after the
parton-shower evolution has completed. Step~\ref{alg:fusing_3} of
the fusing method is therefore implemented on the combined evolution history 
of matrix element and parton shower, i.e.\ the parton-shower history determined
during step~\ref{step:ps_history} of the MEPS@NLO merging, and the subsequent
QCD evolution.

\begin{figure}
    \includegraphics[scale=0.45]{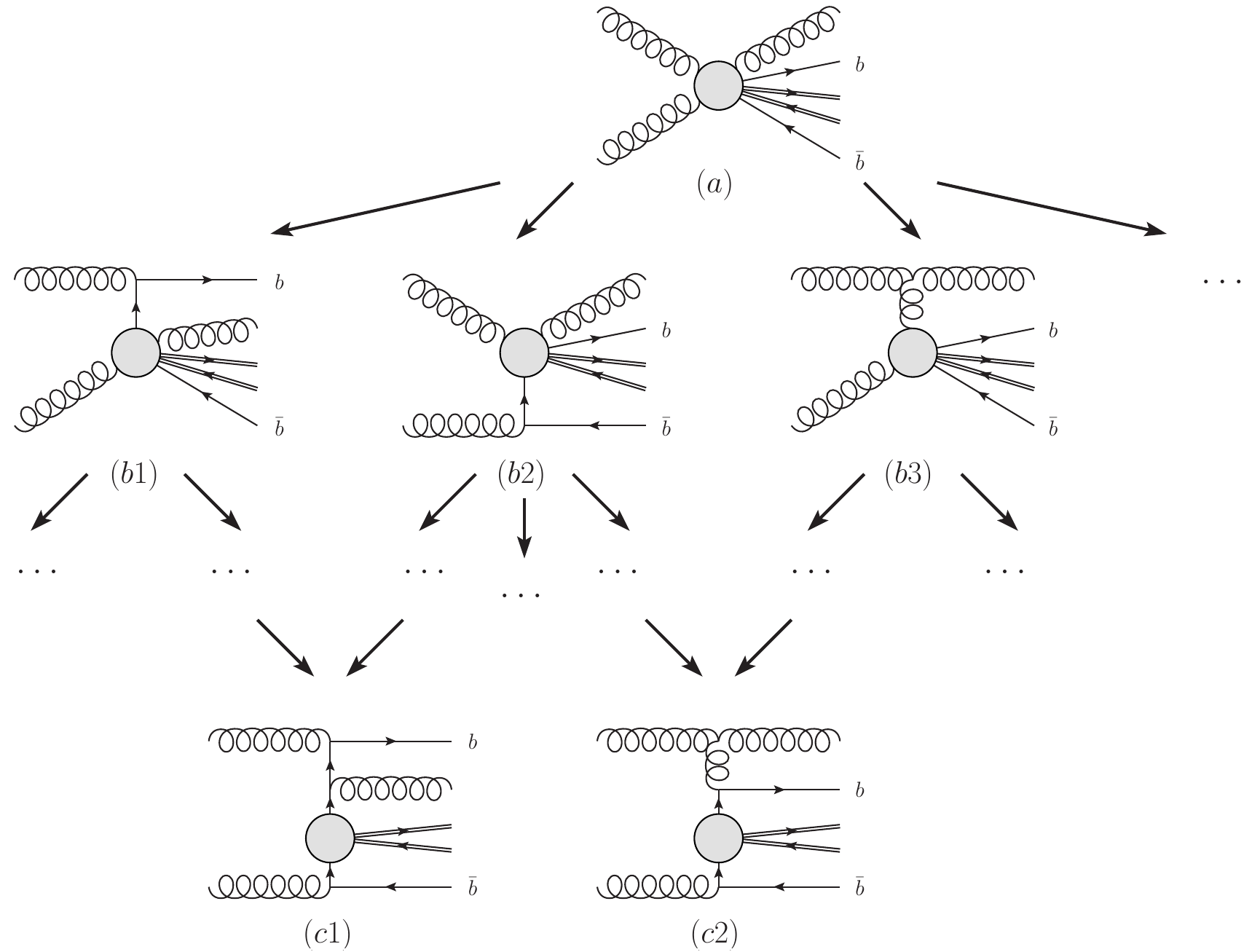}
    \caption{Example parton-shower histories for $gg\to t\bar{t}b\bar{b}g$.
    Top quarks are shown as double lines. Depending on the clustering path,
    configuration (c2) may be identified with a $t\bar{t}b\bar{b}$ topology
    at leading order, while configuration (c1) may not.
    At next-to-leading order, both configuration (c1) and (c2) can be
    identified with a $t\bar{t}b\bar{b}$ topology (see~\cite{Hoche:2019ncc}).
    \label{fig:fusing_examples}}
\end{figure}
Consider the parton-shower histories that are shown in Fig.~\ref{fig:fusing_examples}.
For illustration purposes, we assume that the starting configuration (a) corresponds 
to an event in the inclusive \ttj\ sample, and that the fusing is applied to a
leading-order \ttbb\ calculation. Configuration~(c2) can be reached from configuration~(a)
via~(b2) and~(b3). If the final-state gluon is clustered in the 
first step~(i.e.\ we chose~(b3)), then the configuration~(c2) corresponds
to an event that can be generated in the four flavor-scheme calculation, and it is removed
from the inclusive event sample. If the quark was
clustered first~(i.e.\ we chose~(b2)), then the event is not identified as part
of the four flavor-scheme calculation. It is kept and becomes part of the fragmentation
component in the fusing. The configuration~(c1) is always kept as part
of the fragmentation component. The extension of this algorithm to NLO
precision is described in~\cite{Hoche:2019ncc}.

In order to achieve theoretical consistency between the \ttj\
and the \ttbb\ simulation, the same PDF and strong coupling definition
must be used in both components. Because the heavy-flavor associated
production is a subset of the inclusive result, usage of the five-flavor
PDF and corresponding $\alpha_s$ are mandatory. From the practical 
perspective, five-flavor evolution is important to achieve a Sudakov
suppression of initial-state branchings involving heavy flavors,
and for initial-state $g\rightarrow b\bar{b}$ splittings to be present in parton-shower histories.
Using five-flavor running in a parton-level calculation with massive
\bquark s does however cause a mismatch in the renormalization
of the strong coupling, which needs to be adjusted through
explicit counterterms that depend on the process under consideration.
This approach is identical to the FONLL solution~\cite{Forte:2016sja,Forte:2018ovl}.

It is important to note that even though at NLO QCD the \ttbb\ four-flavor
calculation can model the fragmentation of a high transverse momentum 
light-flavor jet into a pair of heavy flavor quarks, there are qualitatively different
additional contributions from the fragmentation component that can be
substantial in certain regions of phase space.
They correspond to a hard gluon jet fragmenting into a hard
heavy flavor pair, leading to two heavy-flavor pairs in the overall event.
Such configurations have a relatively large probability at high jet transverse
momenta. Modeling them in a fixed-order calculation would require at least
NNLO precision in the direct component, something that is currently far 
out of reach. The effect was explained in detail in~\cite{Cascioli:2013era}.

\begin{figure}
    \includegraphics[scale=0.45]{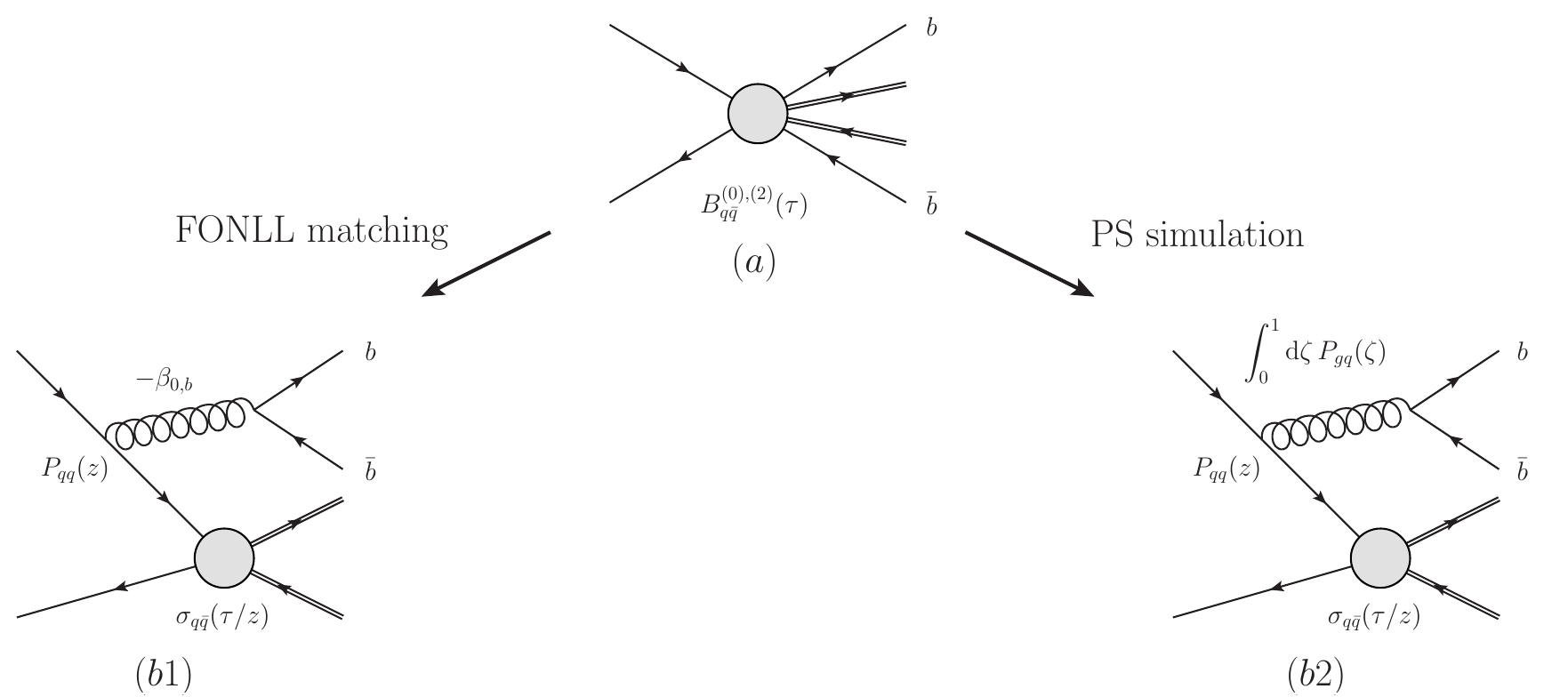}
    \caption{Example for the correspondence between the FONLL matching
      coefficients and the parton-shower expressions. The second order, 
      leading logarithmic coefficient $a_{qq,b}^{(2,2)}(z)=-\beta_{0,b}P_{qq}(z)/2$, with
      $\beta_{0,b}-2/3T_R$~\cite{Buza:1996wv} in (b1) is reproduced~\cite{Hoche:2019ncc}
      by the first-order expansion of the parton-shower result in (b2).
    \label{fig:fusing_vs_hfor}}
\end{figure}
Figure~\ref{fig:fusing_vs_hfor} can be used to understand the correspondence between
the FONLL matching method and the fusing procedure intuitively. Diagram~(a) shows a
complete final-state configuration similar to Fig.~\ref{fig:clustering_examples}
before clustering. This is a typical input for the calculation of the direct component.
Diagrams~(b1) and~(b2) show a typical parton-shower history for diagram~(a).
It is associated with a hard coefficient matrix, $\sigma_{q\bar{q}}$, which will
depend on the scaled center-of-mass energy $\tau=m_{t\bar{t}}/s$ and the $t$-channel
virtuality, which we will call $t$. There are two additional parton splittings in the
history shown in Figs.~\ref{fig:fusing_vs_hfor}~(b1) and~(b2).
They are integrated out between $m_b^2$ and a hard scale $Q^2$, leading to 
enhancements proportional to $\alpha_s^nL^m$, where $L=\ln(Q^2/m_b^2)$. The scale $Q^2$ 
for the diagram at hand is (at leading color) given by $t$. In the FONLL-type matching 
procedure, the coefficient of the leading logarithmic contribution would be given by
the matching term $a_{qq,b}^{(2,2)}$~\cite{Buza:1996wv}, while in a parton-shower simulation
it would correspond to the parton-shower branching probability, given by the leading-order
quark-to-quark splitting function and the integral over the leading-order gluon-to-quark
splitting function. Both results are expansions of the perturbative series to the second order
in the strong coupling, and to leading logarithmic order, and hence they agree. A similar
correspondence will be obtained for all other parton-shower histories~\cite{Hoche:2019ncc}.

We emphasize again that the computation of matching coefficients in the fusing method
is performed algorithmically. The logarithmic precision of our approach is therefore determined
by the logarithmic precision of the parton shower. In \ttj production, we cannot formally claim
more than leading logarithmic precision at leading color. However, we do note that this leads
to a large number of correct matching terms which would not be reproduced in a simple
MC@NLO or POWHEG approach applied to the \ttbb process. Our results are therefore the
best available particle-level prediction for \ttbb\ production to date.

\section{Monte-Carlo Simulation}
\label{sec:simulation}
The fusing algorithm is implemented in the \sherpa event generator~\cite{
Gleisberg:2003xi,Gleisberg:2008ta,Bothmann:2019yzt} in full generality. 
The counterterms and the event veto are realized as userhooks starting with 
version 2.2.12. Tree-level matrix elements and NLO infrared subtraction terms
are calculated by the internal generators Comix~\cite{Gleisberg:2008fv}
and Amegic~\cite{Krauss:2001iv}. Virtual corrections are obtained from
OpenLoops~\cite{Cascioli:2011va}, using CutTools~\cite{Ossola:2007ax}
and OneLoop~\cite{vanHameren:2009dr,vanHameren:2010cp}.

In the hard, perturbative calculation, unstable particles are treated using the complex
mass scheme \cite{Denner:2006ic}. 
The five-flavor PDF set NNPDF~3.0 at NNLO precision with $\alpha_s(M_Z)=0.118$~\cite{NNPDF:2014otw}
is used for both components of the fused prediction. Renormalization, factorization and
resummation scales for both the direct and the fragmentation component are defined
as in the five-flavor \ttj calculation, i.e.\ using the MEPS@NLO merging procedure~\cite{
  Hoeche:2012yf,Gehrmann:2012yg} (cf.\ Sec.~\ref{sec:merging} and Ref.~\cite{Czakon:2019bcq}).
Seven-point factorization and renormalization scale variations are
computed on-the-fly using the techniques in~\cite{Bothmann:2016nao}.
We employ the efficiency improvements for the \sherpa event generator
introduced in~\cite{Danziger:2021xvr}.

The fused predictions will be compared to a merged \ttj five-flavor calculation
and a four-flavor scheme \ttbb computation, both performed at NLO precision in QCD.
While the first uses the same setup as the fused prediction, the latter uses
the four-flavor PDF set of NNPDF~3.0 ~\cite{NNPDF:2014otw} and the corresponding
definition of the strong coupling. The scales of the four-flavor calculation
are defined as~\cite{Jezo:2018yaf}
\begin{equation} \label{eqn:scales}
  \mu_F=\frac{1}{2}\sqrt{\sum{m_{\perp,i}}}\;,\qquad
  \mu_R=\frac{1}{2}\sqrt[4]{\vphantom{\sum}m_{\perp,t}m_{\perp,\bar{t}}\,m_{\perp,b}m_{\perp,\bar{b}}}\;,
  \qquad\text{where}\qquad
  m_\perp=\sqrt{m^2+p_\perp^2}\;.
\end{equation}
Here, the sum runs over all QCD particles, including the top quarks, bottom
quarks, and any additional light partons. Note that these scales are computed
on the partonic final-state before the top decay. They are inspired by the renormalization
scale definitions in the CKKW method~\cite{Catani:2001cc,Amati:1980ch,Catani:1990rr}.
For a typical scale of $b$-jet production of $m_{\perp,b}$ and $m_{\perp,\bar{b}}$,
and a core-process scale given by the geometric average of $m_{\perp,t}$ and $m_{\perp,\bar{t}}$,
the solution of the CKKW constraint $\alpha_s(\mu_R^2)=\alpha_s(m_{\perp,t}m_{\perp,\bar{t}})
\alpha_s(m_{\perp,b}^2)\alpha_s(m_{\perp,\bar{b}}^2)$ at 1-loop order leads to the geometric
average in Eq.~\eqref{eqn:scales} (see also~\cite{Hamilton:2012np}). The additional prefactor
of $1/2$ in Eq.~\eqref{eqn:scales} was determined in~\cite{Buccioni:2019plc} through a detailed
comparison to the NLO prediction for \ttbb+jet.
The parton shower starts from a $2\rightarrow 4$ configuration, which is enforced
in the clustering procedure through the Sherpa configuration parameter
\texttt{CSS\_KMODE=34}~\cite{Cascioli:2013era}.

We use the parton shower developed in~\cite{Schumann:2007mg}, including the changes described
in~\cite{Hoeche:2009xc,Carli:2010cg,Hoeche:2014lxa}. We also extend the new evolution scheme
introduced for improved modeling of Drell-Yan type processes~\cite{ATLAS:2021yza}
to massive particles. The changes to the soft-enhanced terms of
the splitting functions listed in App.~A of~\cite{ATLAS:2021yza} introduce a quark-mass dependent
correction to the flavor-diagonal $\mathbf{K}$ operators of the infrared subtraction
algorithm~\cite{Catani:1996vz}, which is given by
\begin{equation}\label{eq:subscheme_2_ct}
  \mathbf{K}^{ab}(x)\to\mathbf{K}^{ab}(x)
  +\delta^{ab}\sum_i\mathbf{T}_i\mathbf{T}_b\frac{\alpha_s}{2\pi}\left(2\ln\frac{2-x-z_-}{2-x-z_+}
  +2\ln\frac{1-x+u_+}{1-x}\right)\;.
\end{equation}
The boundaries $z_\pm$ and $u_+$ are given by Eqs.~(5.49) and~(5.80)
in~\cite{Catani:2002hc}, respectively. As we modify only the soft-enhanced
part of the splitting functions, we find $z_-=1-u_+$, and $z_+=1$.
Hence, the two additional terms in Eq.~\eqref{eq:subscheme_2_ct} are identical.
Equation~\eqref{eq:subscheme_2_ct} makes the scheme introduced in~\cite{ATLAS:2021yza}
applicable to both massless and massive partons. The changes will be included
in Sherpa version 2.3.1. However, they have little effect on our results,
and we will therefore use the default parton-shower settings in the numerical analysis.

The detailed parameter settings for all calculations are given in App.~\ref{app:runcards}.
We make use of the Rivet analysis toolkit~\cite{Buckley:2010ar}.

\section{Parton-level results with stable top quarks}
\label{sec:validation}

In the following, we present parton-level studies of the fused \ttj $\oplus$ \ttbb\ simulation
of proton-proton  collisions at a center-of-mass energy of 13~TeV.
To gauge the overall impact of the fusing procedure, we compare the prediction
to \ttbb four-flavor and \ttj five-flavor scheme calculations.
In order to identify jets from associated production unambiguously, and to have
access to the proper kinematics of the top quark, the simulation does not include
top quark decays. The event record is parsed at the parton level, before hadronization and
not including the effects of multi-parton interactions.

Jets are reconstructed from final-state partons with a pseudo-rapidity of ${|\eta|}<5$,
using the anti-$k_T$ algorithm~\cite{Cacciari:2008gp} with a radius of $\Delta$R=0.4,
as implemented in FastJet~\cite{Cacciari:2011ma}. Top quarks are excluded from the
jet clustering. Jets are required to have \pt$>$25\,GeV and $|\eta|<2.5$.
They are identified as \bjet s if they contain at least one \bquark.
Jets containing at least two \bquark s are labeled as $B$-jets.

The \ttbar production events are studied in four different phase space regions:
In the inclusive phase space without requiring additional jets or \bjet s (\ttj),
in association with at least one (two) \bjet\  (\oneb, \twob) and in an ``inverse'' region,
which is defined by a light-flavor jet having larger transverse momentum than the \bjet s.
The inclusive selection is used to cross-check the simulation of \ttbar production.
The selection with two \bjet s is the main irreducible background to \tthbb\ and searches
with multiple \bjet s in the final state, which is the focus of our study. 
It is used to highlight the phase-space coverage by the different components
in the fusing  algorithm.
Often, the experimental measurements use regions dominated by \oneb as a control region
for data-driven estimates of \ttbb event rates. To allow for the extrapolation of these
corrections to the \twob\ events, a theoretically consistent treatment of events with different
\bjet\ multiplicity is required. Furthermore, due to the limited capability to separate 
\bjet s from charm  and light jets with high purity, the \oneb{} region usually contains a significant amount
of background from \ttj or \ttc, which needs to be properly integrated into the overall prediction.
 
\begin{figure}
  \subfloat[Number of \bjet s.]{
    \includegraphics[width=0.425\textwidth]{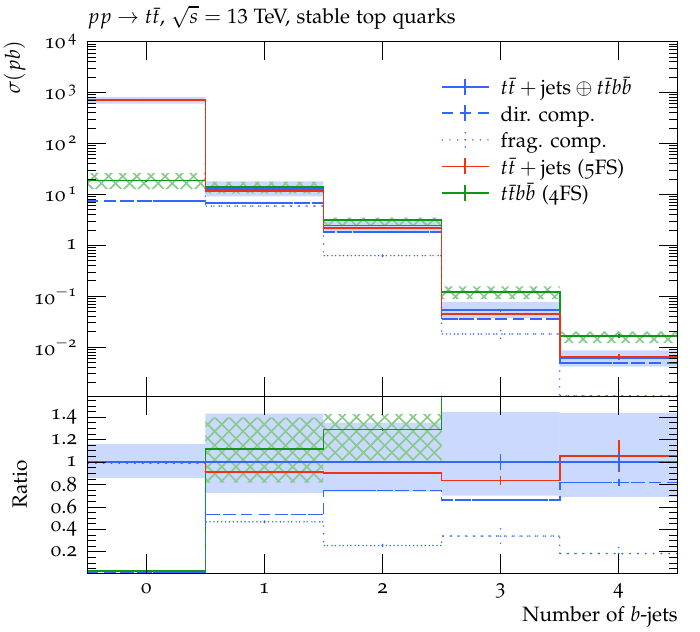}}\hskip 5mm
  \subfloat[Visible transverse energy.]{
    \includegraphics[width=0.425\textwidth]{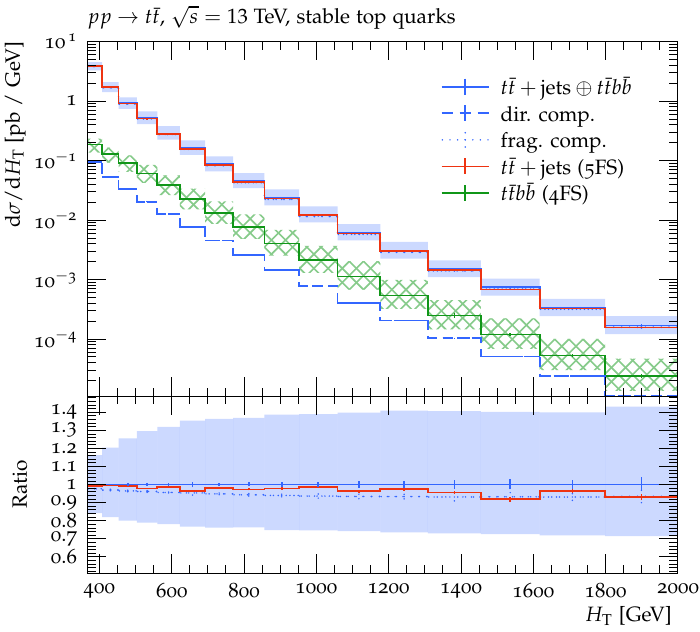}}\\
  \subfloat[Average transverse momentum of top quarks. ]{
    \includegraphics[width=0.425\textwidth]{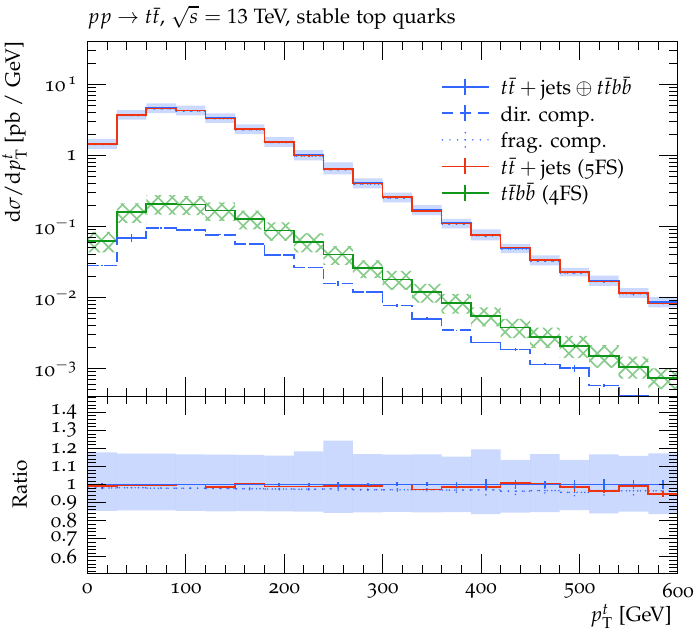}}\hskip 5mm
  \subfloat[Transverse momentum of leading light jet.]{
    \includegraphics[width=0.425\textwidth]{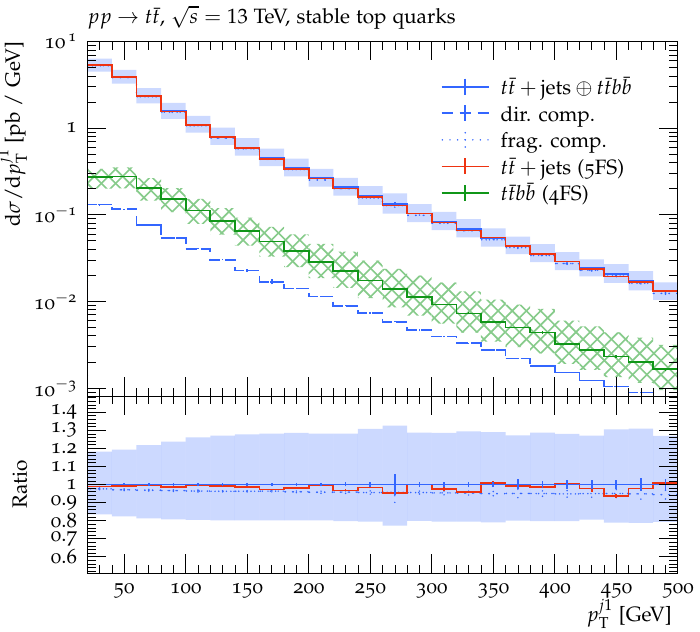}}
  \caption{Predictions of the fusing algorithm in comparison to the  \ttj five-flavor 
    and the \ttbb\ four-flavor scheme calculations in the inclusive \ttbar region,
    using stable top quarks. The blue (solid) and green (hatched) bands are the envelopes
    of the seven-point renormalization and factorization scale variation.
    See the main text for details.}
   \label{fig:stable_inclusive}
\end{figure}
Figure~\ref{fig:stable_inclusive} shows  observables in \ttbar production events 
in the inclusive phase space. The distribution of the \bjet\ multiplicity  illustrates
the key question we are addressing with the fusing algorithm: the \ttj\ five-flavor scheme
prediction deviates significantly from the four-flavor scheme \ttbb result for events
containing  \bjet s, particularly at large \bjet\ multiplicity, indicating the need
for a correction, which is also supported by the disagreement with 
experimental measurements~\cite{ATLAS:2021qou,CMS:2023xjh}.
In addition, the \ttbb calculation does not predict \ttbar events without \bjet s
at an appreciable rate, and can therefore not be extended to the inclusive phase space%
\footnote{Due to the fiducial cuts a few \ttbb\ events still fall into the first
  bin of Fig.\ref{fig:stable_inclusive}~(a)}.   
The fused prediction on the other hand covers the full phase space
with a  smooth transition between regions of different \bjet\ multiplicity:
for events without \bjet s, it consists almost entirely of the fragmentation 
component and therefore agrees very well with the five-flavor scheme result.
This can be seen in the first bin of Fig.~\ref{fig:stable_inclusive}~(a).
With increasing \bjet\ multiplicity, the direct component of the fused prediction
increases from 40\% to ~80\%, leading to differences to  both the \ttj 
and \ttbb result. 
Figs.~\ref{fig:stable_inclusive} (b)-(d) illustrate that the good agreement
at the inclusive level also holds for differential cross sections in the top
transverse momentum, the leading light jet transverse momentum, and the total
visible energy. These distributions are dominated by events without additional
\bjet s, and the direct component does not contribute significantly in any part
of the phase space. The uncertainty on the fused prediction is around 20\% at the
inclusive level and increases to 30-40\% at larger light-jet transverse momentum
and $H_T$.

\begin{figure*}
  \subfloat[Number of light jets.]{
    \includegraphics[width=0.425\textwidth]{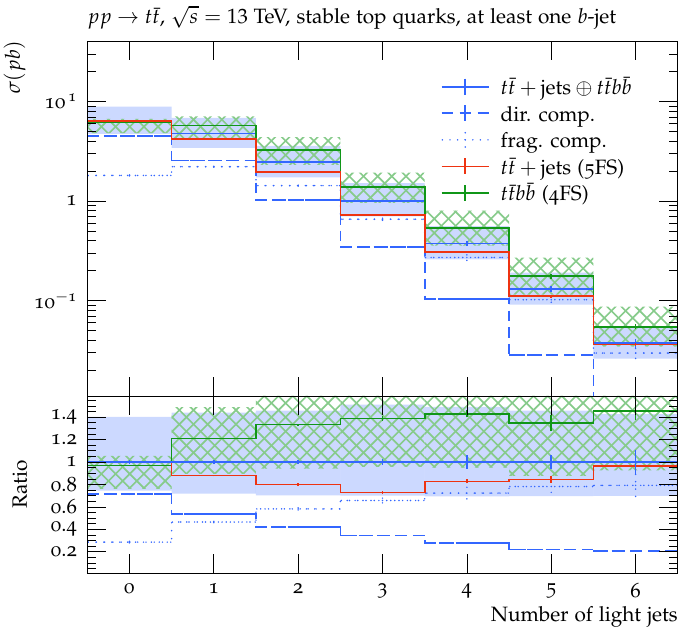}}\hskip 5mm
  \subfloat[Transverse momentum of leading $b$-jet. ]{
    \includegraphics[width=0.425\textwidth]{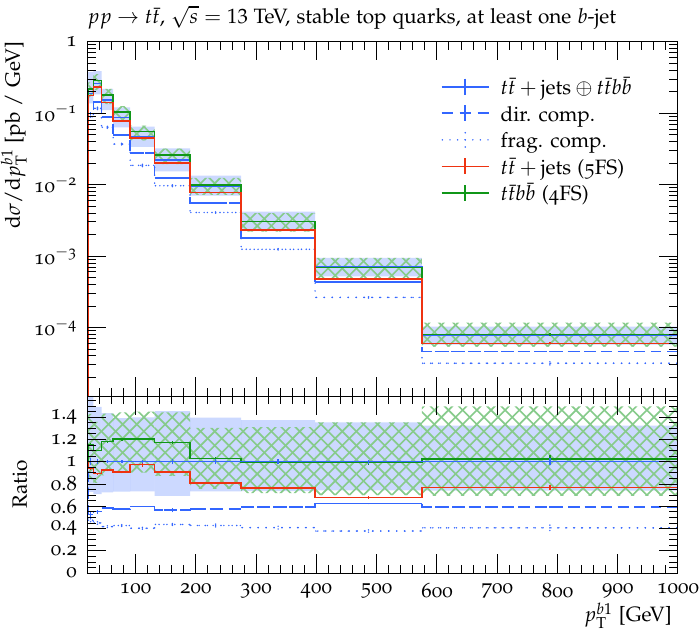}}\\
  \subfloat[Transverse momentum of leading light jet.]{
    \includegraphics[width=0.425\textwidth]{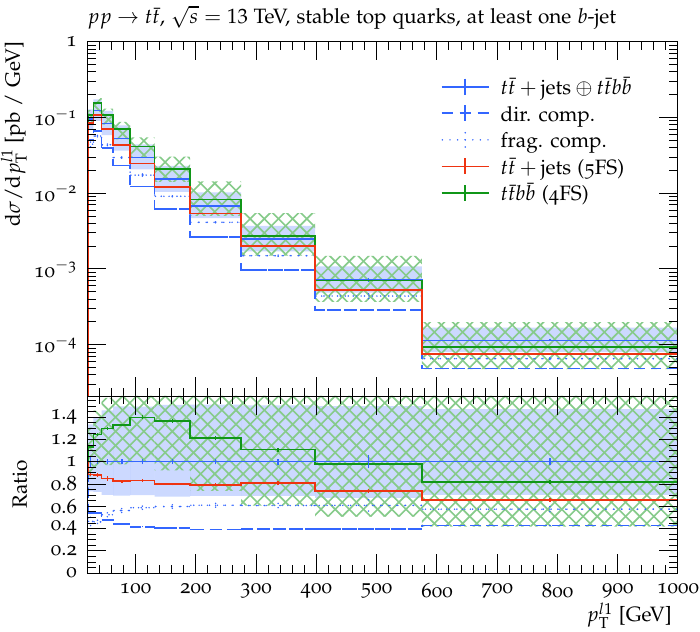}}\hskip 5mm
  \subfloat[Visible transverse energy. ]{
    \includegraphics[width=0.425\textwidth]{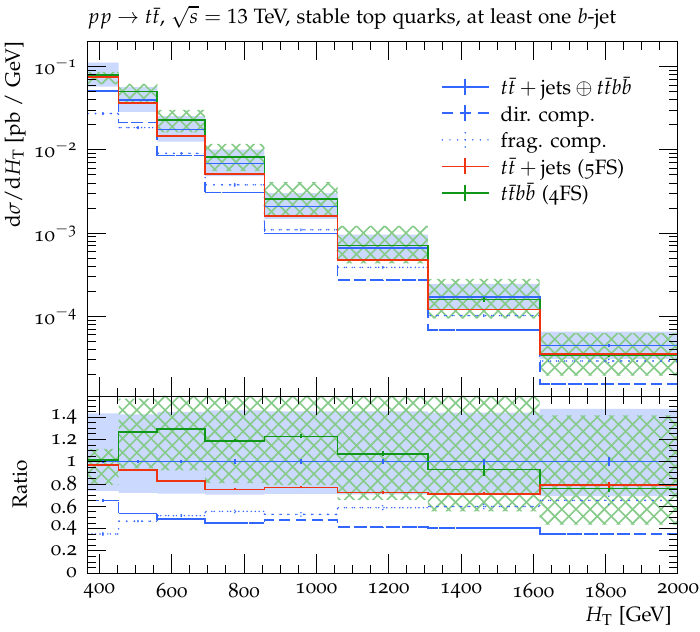}}\\
  \subfloat[Number of $B$-jets.]{
    \includegraphics[width=0.425\textwidth]{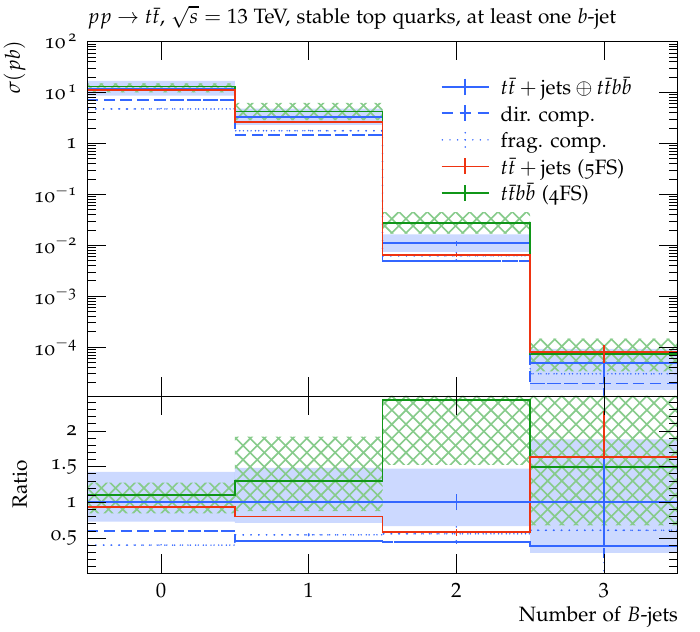}}\hskip 5mm
  \subfloat[Transverse momentum of leading $B$-jet.]{
    \includegraphics[width=0.425\textwidth]{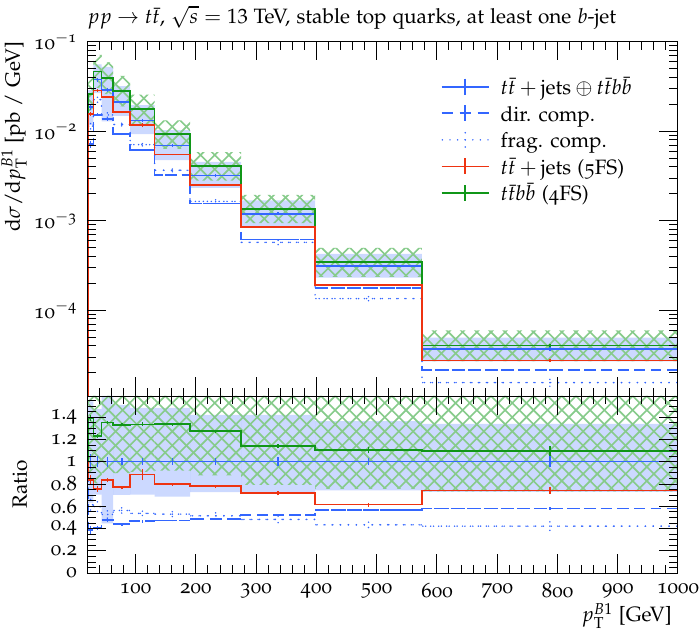}}
  \caption{Predictions of the fusing algorithm in comparison to \ttj five-flavor (5FS) 
    and \ttbb four-flavor (4FS) scheme calculations in the region with at least one \bjet,
    using stable top quarks. The blue (solid) and green (hatched) bands are the envelopes
    of the seven-point renormalization and factorization scale variation.
    Uncertainty bands are cut if they exceed the axis range in the ratio panel.
    See the main text for details.}
\label{fig:stable:atleastoneb}
\end{figure*}
Figure~\ref{fig:stable:atleastoneb} shows observables in the \oneb\ region.
The exclusive light-jet multiplicity shows that the phase-space region without additional
QCD activity is dominated by the direct contribution and the contribution of the fragmentation
component increases to 80\% with increasing number of light jets.  It is interesting to observe in particular
that the four-flavor and five-flavor calculations agree if no additional light jet is present,
while the four-flavor cross section exceeds the five-flavor cross section in the case of few
additional light jets.
We remind the reader that the fused event sample presented here consists of a next-to-leading order
precise computation for \ttbb and a \ttj sample with \ttbar+0~jet computed at next-to-leading order,
and \ttbar+1-4~jets computed at leading order in QCD.
The uncertainty of the fused prediction is larger than the \ttbb\ four-flavor scheme uncertainty
partly because of the less precise \ttbar+2~jet input calculation in the fragmentation component.
This uncertainty can be reduced through substituting the \ttbar+1~jet and \ttbar+2~jet inputs by
next-to-leading order predictions. It was observed in~\cite{Hoeche:2014qda} that such a change 
leaves the central value of the light-jet simulations mostly unaffected. In order to reduce our
computing budget, we therefore postpone this improvement to a future study.

\begin{figure}
  \subfloat[Visible energy distribution, inverse selection.]{
    \includegraphics[width=0.425\textwidth]{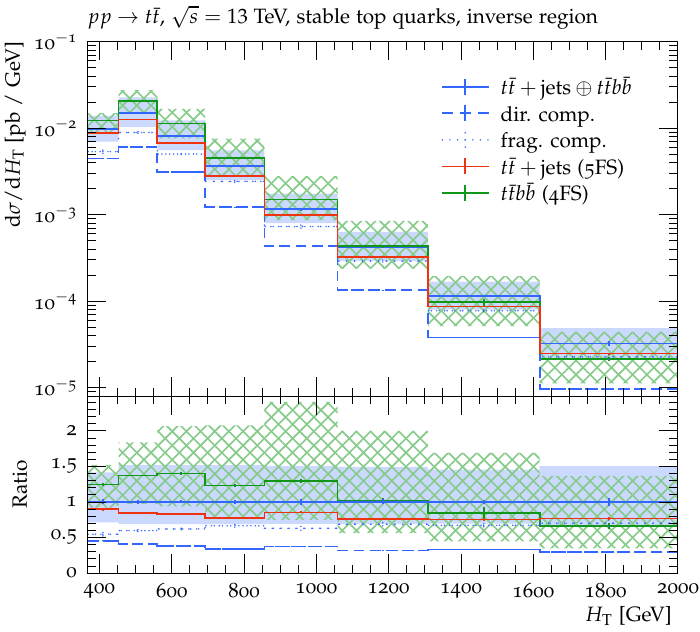}}\hskip 5mm
  \subfloat[Invariant mass of leading \bjet s, inverse selection.]{
    \includegraphics[width=0.425\textwidth]{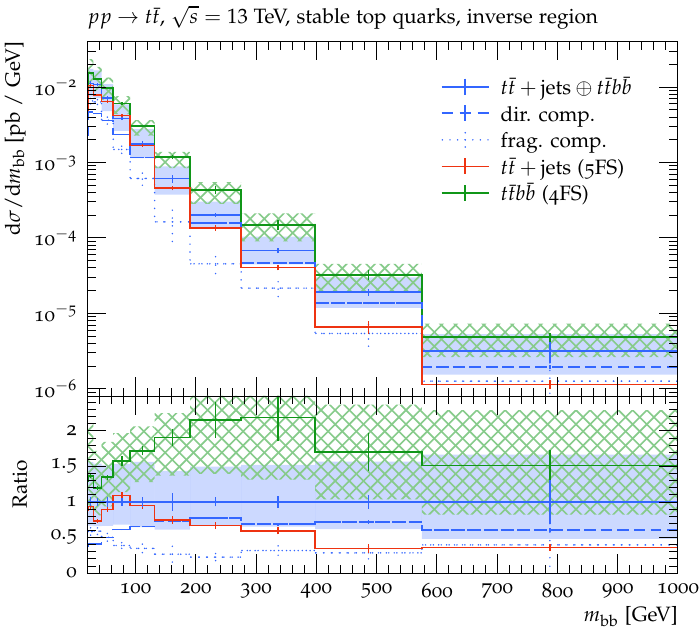}}
  \caption{Predictions of the fusing algorithm in comparison to \ttj five-flavor 
    and the \ttbb four-flavor scheme calculations in the ``inverse'' selection,
    using stable top quarks. See the main text for details.
    \label{fig:inverse_region}}
\end{figure}

The transverse momenta of the leading\footnote{Jets and \bjet s are ordered in \pt.} \bjet\ 
and the leading light jet show significantly different behavior 
between the fused and the other calculations, with the fusing interpolating
between the four-flavor and five-flavor prediction. Similar features also appear in the scalar sum
of the transverse momenta of all jets, $H_T$. While the contribution of the direct component is 
approximately constant at 60\% in the leading \bjet\ \pt, it drops to about 40\% at moderately large
leading light jet \pt ($\geq$ 100\,GeV), and related to this at high $H_T$ ($H_T$ $\geq$ 600\,GeV).
The smaller contribution of the direct component at large light jet transverse momentum is caused
by the Sudakov suppression effects discussed in Sec.~\ref{sec:fusing}.
It will be interesting to investigate the region where the \bjet\ is not the hardest QCD object
(apart from the top), but where it has smaller transverse momentum than the light jet. This
``inverse'' region is studied in Fig.\,\ref{fig:inverse_region} and will be discussed later. 

We also study the rate and kinematic distribution of \bjet s containing at least two $b$-quarks,
labeled as $B$-jets. These type of jets appear when the opening angle in a gluon to $b$-quark pair
splitting is small, which may happen in particular in double splitting events~\cite{Cascioli:2011va}.
A similar amount of events without $B$-jet is found for all calculations. However, the \ttbb\ four-flavor scheme  calculation
predicts a much higher rate with up to 2.5 times the rate of the fused prediction for events with
two $B$-jets and an increased uncertainty on the prediction. The five-flavor \ttj\ calculation
predicts significantly less events with $B$-jets, again due to the Sudakov suppression discussed
in Sec.~\ref{sec:fusing}, which is absent in the four-flavor scheme result. Similar to the \bjet\ 
transverse momentum distribution, the fused prediction lies between the four-flavor and the five-flavor
scheme result. However, at low $B$-jet transverse momentum, the contribution from the fragmentation
component is substantially larger, indicating again a sizable Sudakov suppression of the direct component.

\begin{figure*}
 \hskip 5mm
  \subfloat[Transverse momentum of subleading $b$-jet. ]{
    \includegraphics[width=0.425\textwidth]{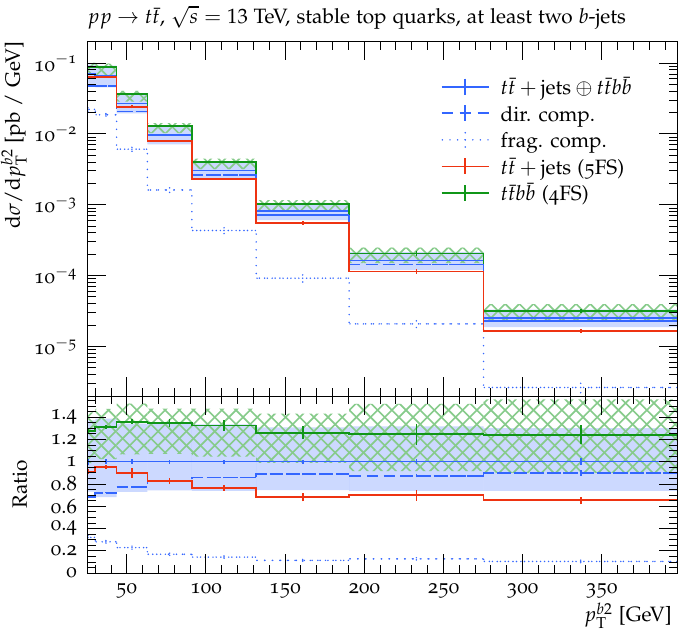}}\hskip 5mm
  \subfloat[Transverse momentum of leading $B$-jet.]{
    \includegraphics[width=0.425\textwidth]{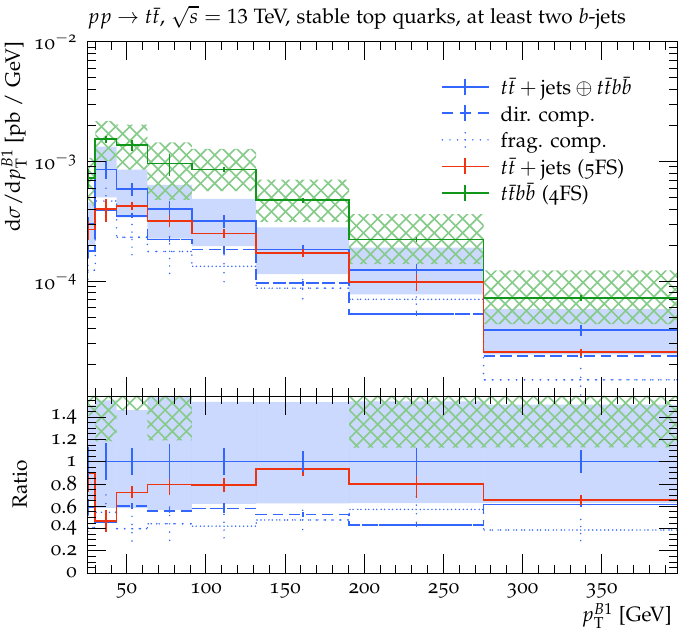}}\\
  \subfloat[Transverse momentum of leading light jet.]{
    \includegraphics[width=0.425\textwidth]{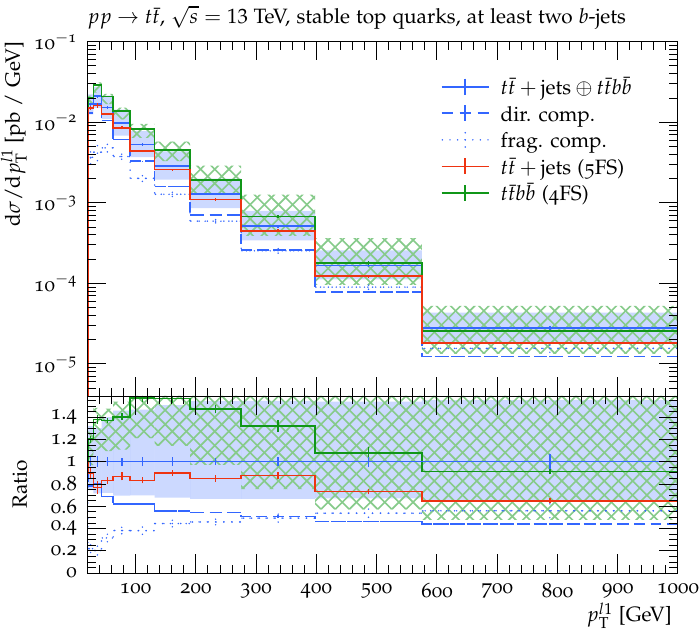}}\hskip 5mm
  \subfloat[Visible transverse energy.]{
    \includegraphics[width=0.425\textwidth]{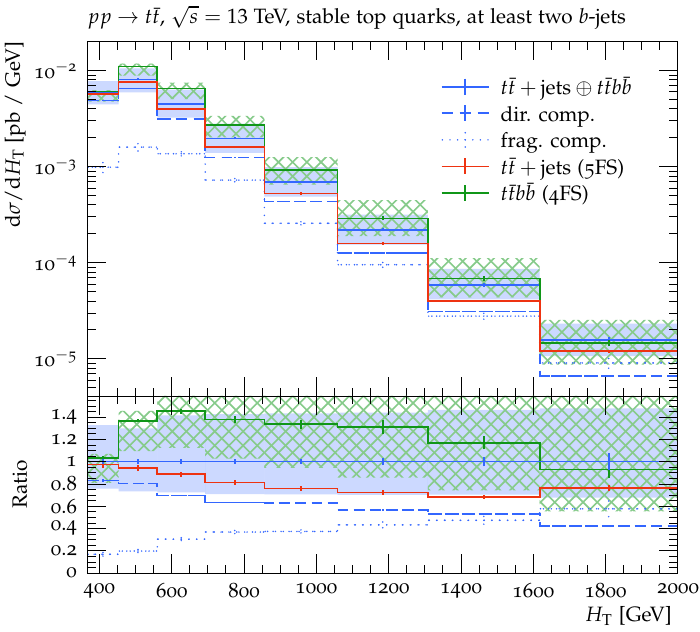}}\\
  \subfloat[$\Delta R$ separation between leading \bjet s.]{
    \includegraphics[width=0.425\textwidth]{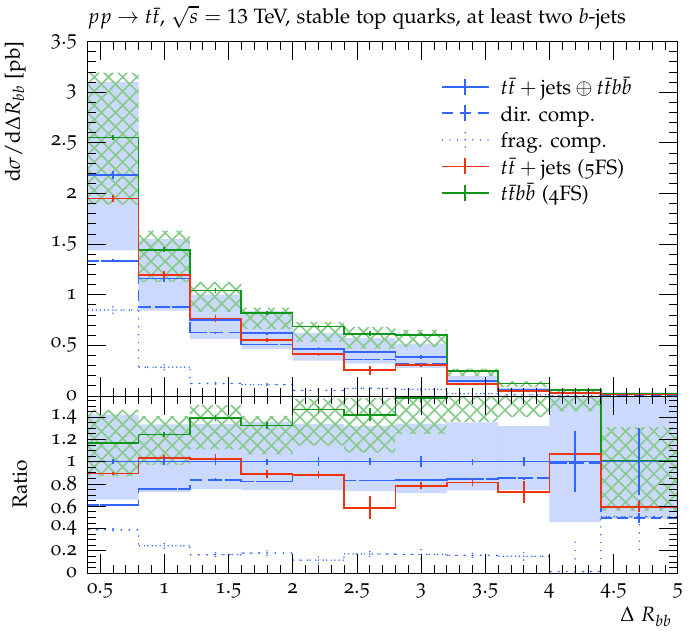}}\hskip 5mm
  \subfloat[Invariant mass of leading \bjet s.]{
    \includegraphics[width=0.425\textwidth]{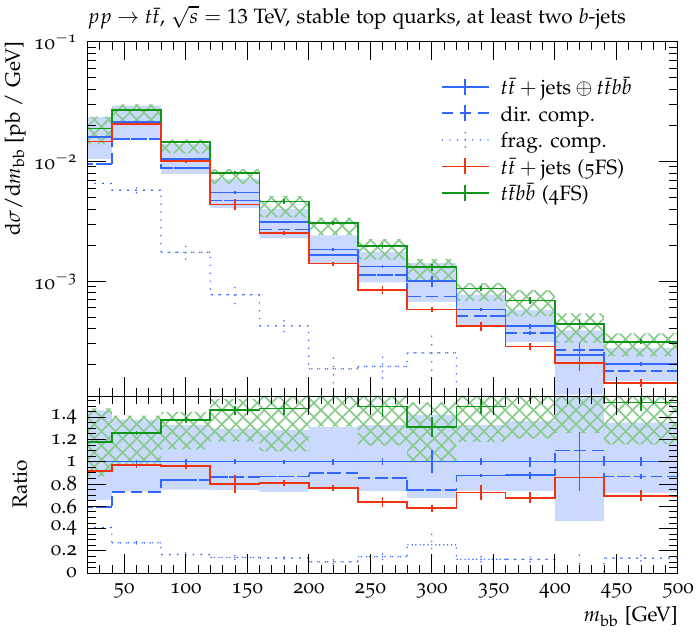}}
  \caption{Predictions of the fusing algorithm in comparison to \ttj five-flavor (5FS) 
    and the \ttbb four-flavor (4FS) scheme  calculations in the region with at least two  \bjet s,
    using stable top quarks. The blue (solid) and green (hatched) bands are the envelopes
    of the seven-point renormalization and factorization scale variation.  
    See the main text for details.}
   \label{fig:stable:atleasttwobs}
\end{figure*}
Figure~\ref{fig:stable:atleasttwobs} shows observables in the \twob region.
The transverse momentum spectra of the leading jet and sub-leading \bjet\ are similar
in shape between the four-flavor scheme prediction and the fused result at large
transverse momentum, but there is a notable change in shape at low \pt for the
leading jet. It is interesting that in this event selection, we find a larger difference
between the five-flavor scheme prediction and the fused result, with the fused prediction
interpolating between \ttj and \ttbb. A sizable suppression is also visible in the invariant
mass between the leading \bjet s at high di-\bjet\ invariant mass.
These effects can be attributed to the Sudakov suppression at high scales, which is
unaccounted for in the four-flavor scheme. The fused result for the sub-leading \bjet\
\pt and the di-\bjet\ invariant mass is dominated by the direct component, as expected.

Figure~\ref{fig:inverse_region} shows the visible transverse energy and the
invariant mass distribution of the closest \bjet s in the ``inverse'' selection.
The four-flavor scheme prediction for the invariant mass deviates substantially 
from the five-flavor scheme result and the fused prediction in this region of 
phase space, and the direct component is only about 60-70\% of the fused prediction
at large $m_{bb}$. The four-flavor scheme predicts relatively large 
parametric uncertainties for both observables.

\section{Particle-level results including top-quark decays}
\label{sec:results}
\begin{figure*}
  \subfloat[Average $\Delta R_{bb}$; $3b$, $\ge 5$jet selection.]{
    \includegraphics[width=0.425\textwidth]{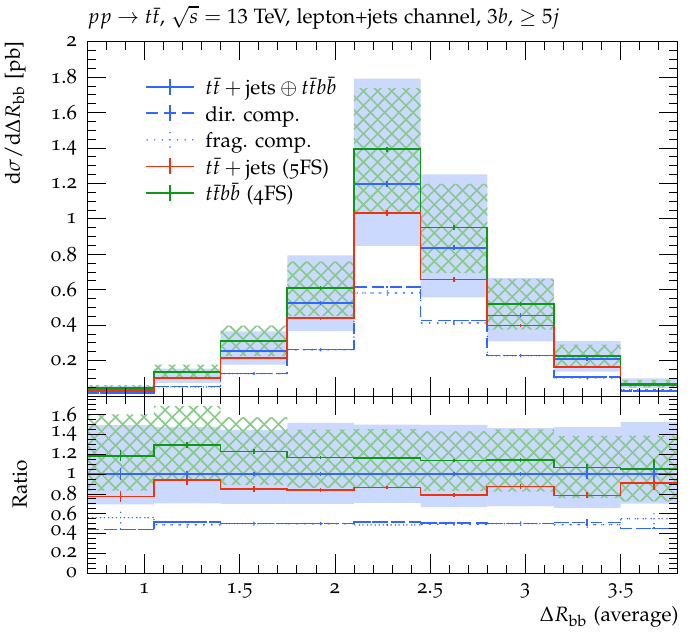}}\hskip 5mm
  \subfloat[Average $\Delta R_{bb}$; in $\ge 4b$, $\ge 6$jet selection.]{
    \includegraphics[width=0.425\textwidth]{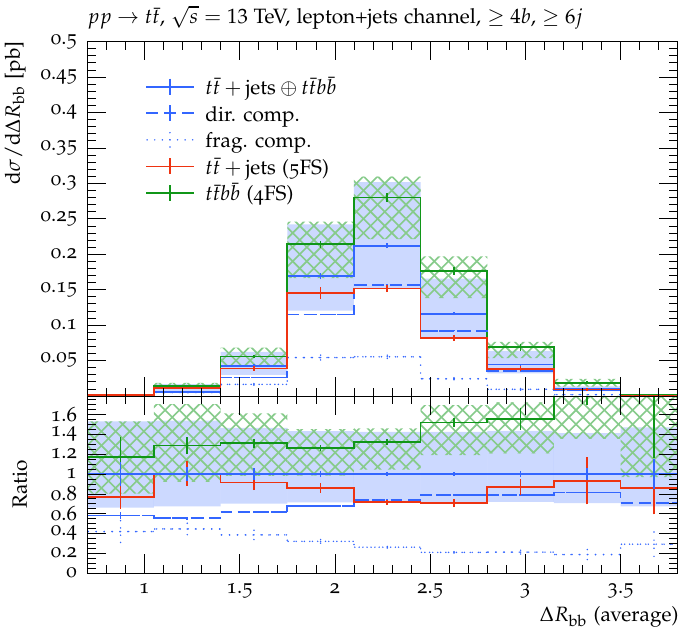}}\\
  \subfloat[$\Delta R_{bb}$ of closest pair; $3b$, $\ge 5$jet selection.]{
    \includegraphics[width=0.425\textwidth]{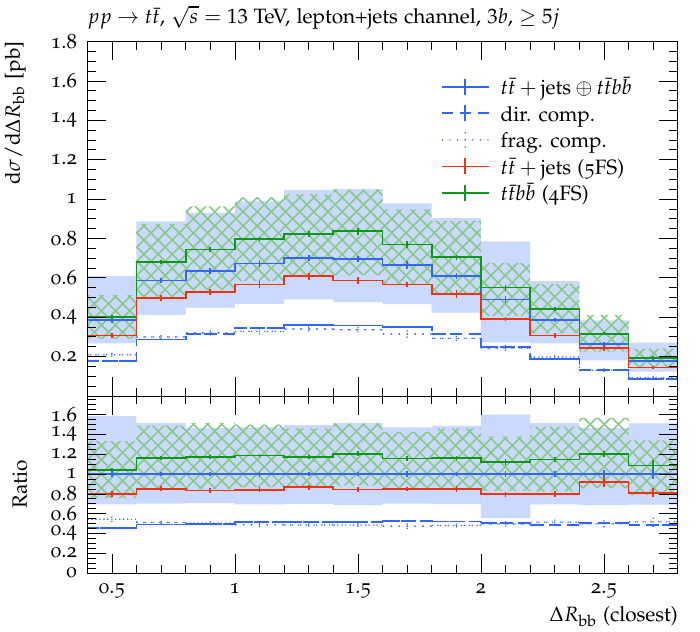}}\hskip 5mm
  \subfloat[$\Delta R_{bb}$ of closest pair; $\ge 4b$, $\ge 6$jet selection.]{
    \includegraphics[width=0.425\textwidth]{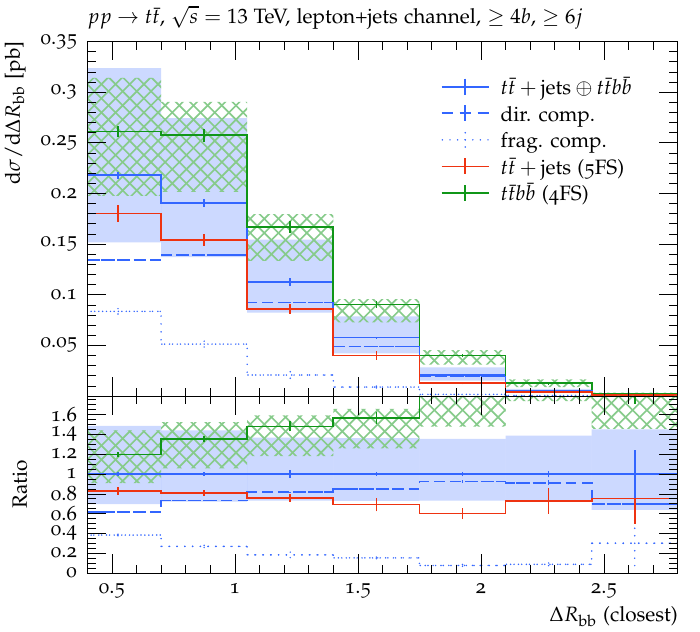}}
 \caption{Predictions of the fusing algorithm in comparison to five-flavor \ttj\
    and four-flavor scheme \ttbb\ calculations using \ttbar\ events in the single lepton decay channel. 
    The blue (solid) and green (hatched) bands are the envelopes
    of the seven-point renormalization and factorization scale variation.
    See the main text for details.}
   \label{fig:decayed:ttbb1}
\end{figure*}
\begin{figure*}
  \subfloat[Visible transverse energy; $3b$, $\ge 5$jet selection.]{
    \includegraphics[width=0.425\textwidth]{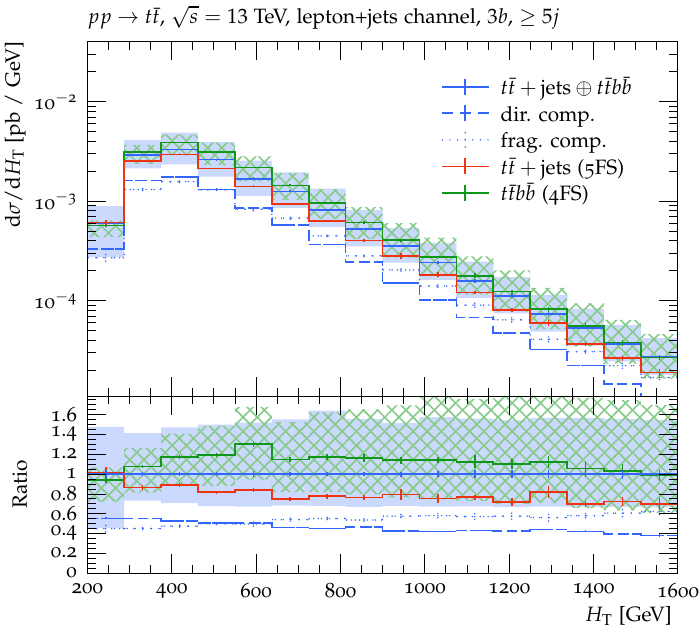}}\hskip 5mm
  \subfloat[Visible transverse energy; $\ge 4b$, $\ge 6$jet selection.]{
    \includegraphics[width=0.425\textwidth]{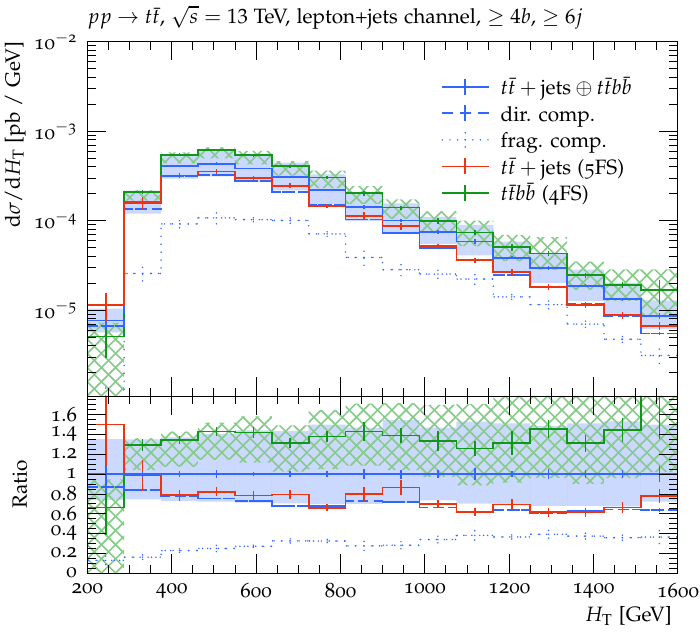}}\\
  \subfloat[Light-jet transverse momentum; $3b$, $\ge 5$jet selection.]{
    \includegraphics[width=0.425\textwidth]{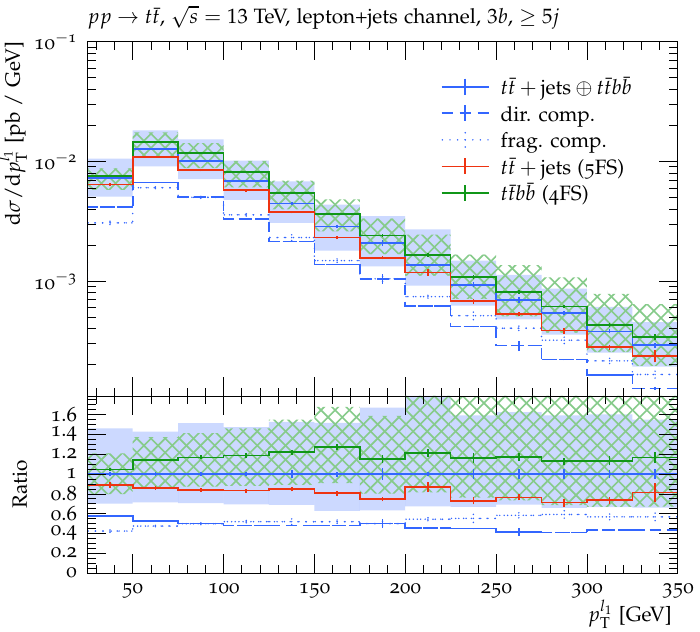}}\hskip 5mm
  \subfloat[Light-jet transverse momentum; $\ge 4b$, $\ge 6$jet selection.]{
    \includegraphics[width=0.425\textwidth]{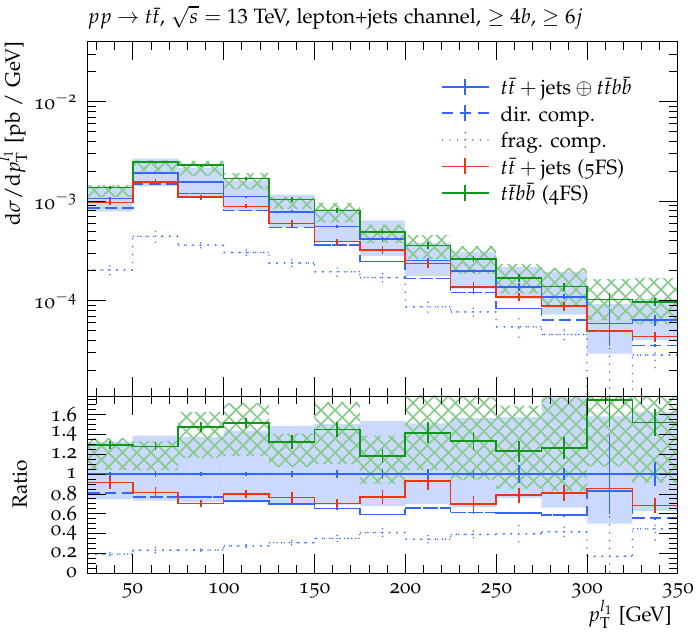}}
  \caption{Predictions of the fusing algorithm in comparison to five-flavor \ttj
    and four-flavor scheme \ttbb\ calculations using top pair production events in the single-lepton channel. 
    The blue (solid) and green (hatched) bands are the envelopes
    of the seven-point renormalization and factorization scale variation.
    See the main text for details.
   \label{fig:decayed:ttbb2}}
\end{figure*}
In Figs.~\ref{fig:decayed:ttbb1} and~\ref{fig:decayed:ttbb2} we investigate
the impact of the fusing on the observables in a Monte-Carlo analysis 
based on the analysis in \cite{Ferencz:2023fso}  that is designed to cover
a fiducial phase space similar to the experimental measurements. 
To obtain realistic final states, top-quark decays and subsequent $W$ boson decays
are simulated by the decay module of \sherpa~\cite{Laubrich:2006aa,Siegert:2006xx},
taking into account spin correlations between the hard process
and the decays~\cite{Richardson:2001df}.

The object definition and event selection applied in this  study is defined at particle level.
All objects are defined using stable final-state particles with a mean lifetime of $\tau > 3\cdot 10^{-11}$s.
Jets are reconstructed from all stable final-state particles (but excluding leptons and neutrinos
from the top quark decay chain) using  the anti-$k_{\mathrm{t}}$ 
jet algorithm~\cite{Cacciari:2008gp,Cacciari:2011ma} with a radius parameter of $R = 0.4$.
Jets which contain at least one ghost-associated~\cite{Cacciari:2008gn} $B$-hadron 
with \pt$>5$\,GeV are defined as \bjet s, all other jets are considered ``light'' jets.
The four-momentum of the bare leptons from top quark decay are modified (``dressed'') 
by adding the four-momenta of all radiated photons within a cone of size $\Delta R=0.1$.
All objects are considered within pseudo-rapidity $|\eta|<2.5$ and with \pt$>27$\,GeV
for leptons and  \pt$>25$  for jets and \bjet s. Leptons are removed if they are separated
from a jet by less than $\Delta R=0.4$, where $\Delta R = \sqrt{(\Delta \eta )^2 + (\Delta \phi)^2}$.
Events are selected with at least four \bjet s, with exactly one lepton and at least six jets.

The $\Delta R$ separations in Fig.~\ref{fig:decayed:ttbb1} show good agreement
between the four-flavor scheme result and the fused prediction in the selection with
three \bjet s and at least five jets. Here, the direct and fragmentation component
contribute roughly equally to the observable, and they have the same shape.
In the region with four \bjet s and at least six jets, the direct component is more
important, but the fragmentation component still contributes at about 40\% at small $\Delta R$.
This is somewhat expected, despite the fact that for large $\Delta R$ the final state can be 
modeled by the four-flavor result. The larger suppression of the direct component
at smaller $\Delta R$ can be attributed to the Sudakov suppression of \bjet\ production
that is not well modeled in the fixed-order four-flavor scheme calculation.

Figure.~\ref{fig:decayed:ttbb2} shows good agreement between the fused and the 
four-flavor scheme results in most regions. The shape of the distributions is
more similar in the $\ge 4b$, $\ge6j$ selection, for which the four-flavor scheme
calculation is more predictive. In this selection, the direct component contributes
up to $\sim80$\% to the fused cross section, while in the $3b$, $\ge 5j$ selection
it contributes about 50\%.

\begin{figure*}
  \subfloat[Fiducial cross sections.]{
    \includegraphics[width=0.425\textwidth]{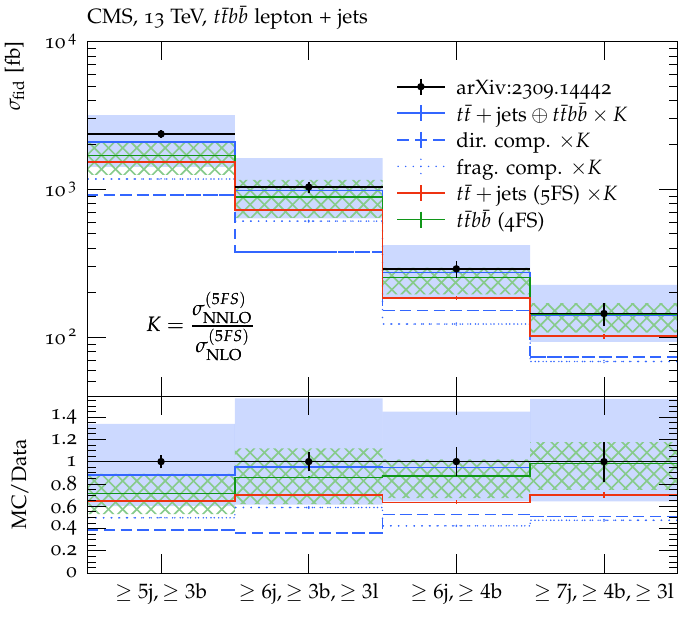}}\hskip 5mm
  \subfloat[Number of jets.]{
    \includegraphics[width=0.425\textwidth]{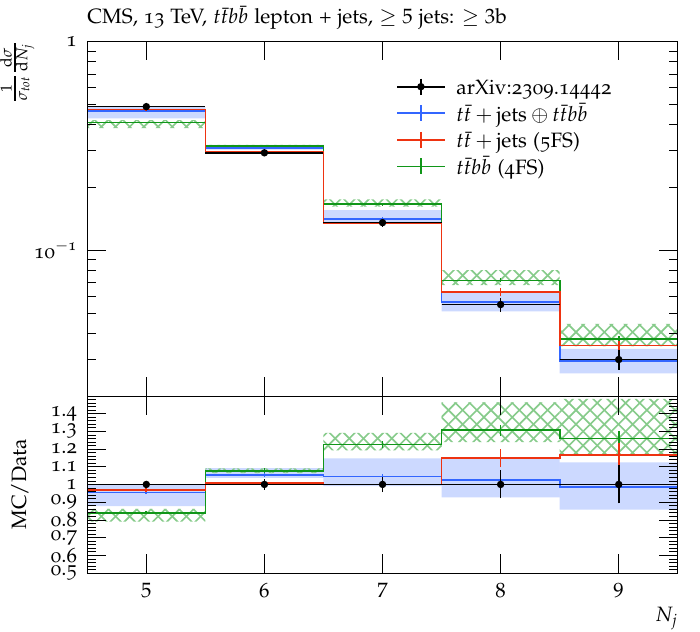}}\\
  \subfloat[Average angular separation between \bjet s.]{
    \includegraphics[width=0.425\textwidth]{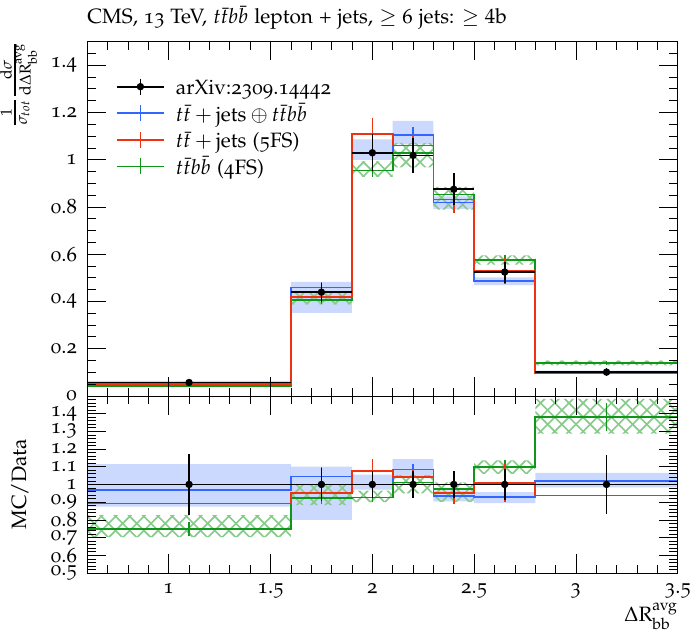}}\hskip 5mm
  \subfloat[Visible transverse energy of \bjet s.]{
    \includegraphics[width=0.425\textwidth]{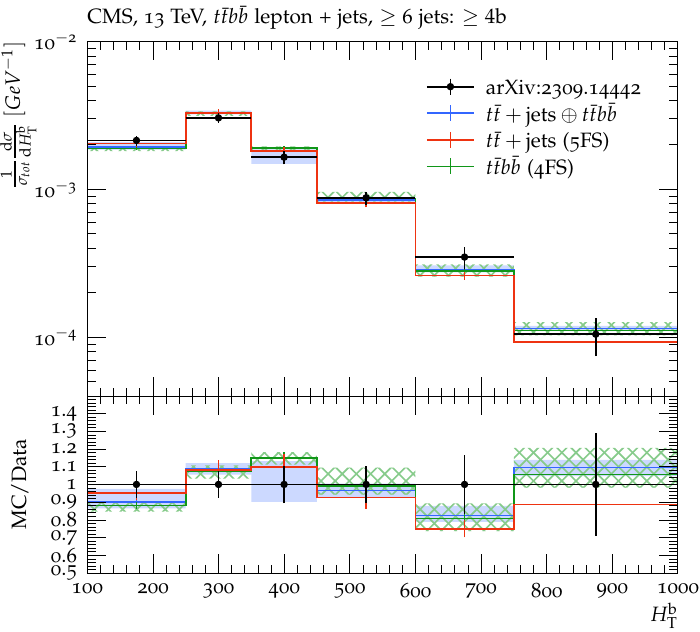}}\\
  \subfloat[\bjet\ transverse momentum.]{
    \includegraphics[width=0.425\textwidth]{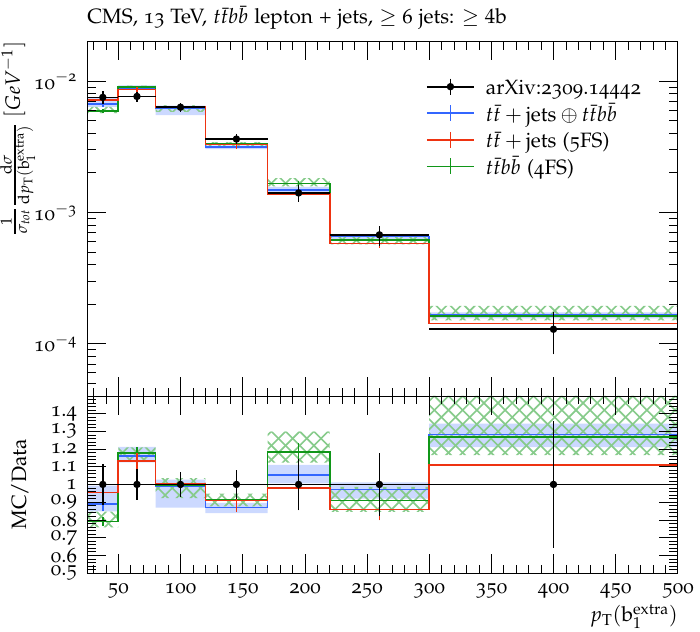}}\hskip 5mm
  \subfloat[Total visible transverse energy.]{
    \includegraphics[width=0.425\textwidth]{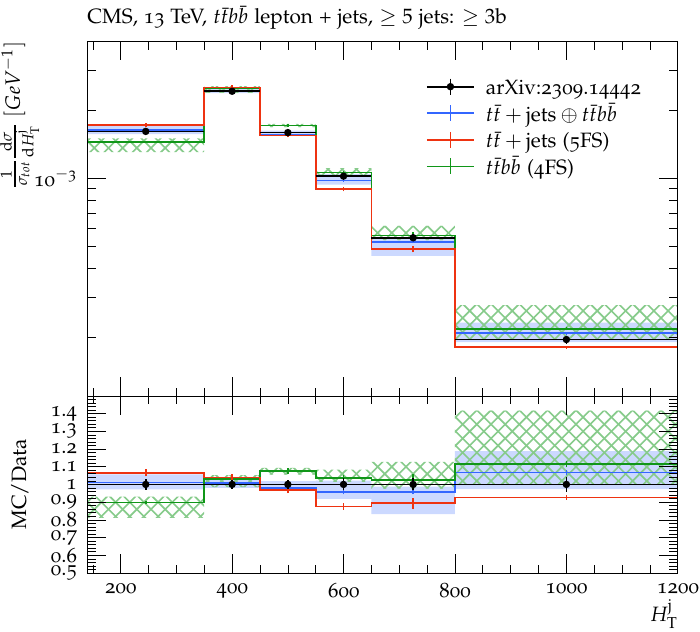}}
  \caption{Comparison of the fused predictions to CMS data from~\cite{CMS:2023xjh}.
    The blue (solid) and green (hatched) bands are the envelopes
    of the seven-point renormalization and factorization scale variation.}
   \label{fig:decayed:CMS}
\end{figure*}
Figure.~\ref{fig:decayed:CMS} shows Monte-Carlo predictions from Sherpa 
in comparison to the CMS measurements in~\cite{CMS:2023xjh}.
The five-flavor scheme, fragmentation component, direct component and fused results
are all scaled by the ratio between the NLO and NNLO inclusive $t\bar{t}$ production
cross section~\cite{Czakon:2013goa,Czakon:2011xx,TopWGPage}. 
We find good agreement between the fused prediction and the experimental data
for the fiducial cross sections, with a particularly significant improvement
over the four-flavor scheme calculation in the $\ge 5j, \ge3b$ region.
It is interesting to note that the direct and fragmentation components contribute
roughly equally to all regions, even in the $\ge 4b$ selection. This indicates that
the proper modeling of the fragmentation component is highly relevant for
typical experimental selection cuts. The increased uncertainty of the fused 
prediction is a consequence of the large contribution of the fragmentation
component. A reduction of this uncertainty can be expected when the $t\bar{t}$+2j
component of the fragmentation component is included at NLO precision.
As already observed in Fig.~\ref{fig:decayed:ttbb1}, the $\Delta R$ separation
between the two leading \bjet s also shows a marked change in shape,
with the fused result in better agreement with experimental data.
The total transverse energy of \bjet s and the \bjet\ transverse momentum in the
$4b$ region show a mild change between the four-flavor scheme and the fused result,
indicating a proper physics modeling by the fusing procedure.
The total transverse energy in the $3b$ selection is mildly improved over the
four-flavor scheme result.

\section{Conclusions}
\label{sec:conclusions}
With increasing luminosity  at the LHC, rare processes with multiple \bjet s 
in the final state are within reach of experimental measurements. The analyses do,
however suffer from background of top pair production in association with light
and heavy flavor jets, which is in many cases irreducible and a source of significant
uncertainties. To address this problem, we have presented a consistent scheme for
matching a fully differential \ttbb\ NLO calculation in the four-flavor scheme
to an inclusive MEPS@NLO merged simulation of \ttj\ production. This approach is
dubbed ``fusing''. We find good agreement between the four-flavor scheme predictions
and the fused result in the region of phase space where two hard \bjet s are observed,
and an increased cross section in regions with only three required \bjet s.

In the future, our technique can be utilized to reduce systematic theory uncertainties
in measurements with multiple \bjet s in the final state -- such as the \tthbb process -- 
through a calibration of the light-jet production rate by means of the inclusive
\ttj\ process. We defer the detailed study of this method to a future publication.

\section{Acknowledgments}
We thank Johannes Krause and Chris Pollard for collaboration in the early stages
of this project, and in particular for contributing the initial version of the
Rivet analysis routine~\cite{ttbbrivet}.
This work was supported by the Fermi National Accelerator Laboratory (Fermilab),
a U.S. Department of Energy, Office of Science, HEP User Facility.
Fermilab is managed by Fermi Research Alliance, LLC (FRA),
acting under Contract No. DE--AC02--07CH11359.
This research used the Fermilab Wilson Institutional Cluster for code development,
testing and validation. We are grateful to James Simone for his support.\\

The material is available under an Open Access CC-BY 4.0 license

\clearpage
\appendix
\section{Sherpa setups}
\label{app:runcards}
Listing~\ref{runcard-4f} shows the Sherpa runcard used for the four-flavor scheme
prediction. We do not include settings related to the top decay handling, which differ
between the stable and the decayed top analysis and are irrelevant for the study
presented in this manuscript.
\begin{lstlisting}[label=runcard-4f,captionpos=b,
  caption={\sherpa runcard for the four-flavor scheme calculation.},
  linerange={1-28,42-51}]
(run){
  MI_HANDLER None;
  ME_SIGNAL_GENERATOR Comix Amegic LOOPGEN;
  EVENT_GENERATION_MODE Weighted;
  LOOPGEN:=OpenLoops;
  # collider setup
  BEAM_1 2212; BEAM_ENERGY_1 6500;
  BEAM_2 2212; BEAM_ENERGY_2 6500;
  # scales and PDFs
  SCALE_VARIATIONS 0.25,0.25 0.25,1. 1.,0.25 1.,1. 1.,4. 4.,1. 4.,4.;
  SCALES VAR{H_TM2/4}{0.25*P_TM2}{H_TM2/4};
  CORE_SCALE VAR{MU_F2}{MU_R2}{H_TM2/4};
  EXCLUSIVE_CLUSTER_MODE 1;
  PDF_LIBRARY LHAPDFSherpa;
  PDF_SET NNPDF30_nnlo_as_0118_nf_4;
  USE_PDF_ALPHAS 1;
  # parton shower and matching
  NLO_CSS_PSMODE 1;
  CSS_RESPECT_Q2 1;
  CSS_KMODE 34;
  CSS_SCALE_SCHEME 20;
  CSS_EVOLUTION_SCHEME 30;
  CSS_IS_AS_FAC 1;
  # model and flavor parameters
  EW_SCHEME 3; GF 1.1663787e-5;
  MASS[5] 4.75; MASSIVE[5] 1;
  MASS[6] 172.5; WIDTH[6] 0; STABLE[6] 0;
  MASS[24] 80.385; STABLE[24] 0;
}(run)
(processes){
  Process 93 93 -> 6 -6 5 -5
  Order (*,0)
  NLO_QCD_Mode MC@NLO
  Loop_Generator LOOPGEN
  ME_Generator Amegic
  RS_ME_Generator Comix
  End process
}(processes)
\end{lstlisting}
\clearpage
Listing~\ref{runcard-direct} shows the Sherpa runcard used for the direct component.
We do not include settings related to the top decay handling, which differ
between the stable and the decayed top analysis and are irrelevant for the
fusing procedure studied here.
\begin{lstlisting}[label=runcard-direct,captionpos=b,
  caption={\sherpa runcard for the direct component of the fusing.},
  linerange={1-31,45-56}]
(run){
  MI_HANDLER None;
  ME_SIGNAL_GENERATOR Comix Amegic LOOPGEN;
  EVENT_GENERATION_MODE Weighted;
  LOOPGEN:=OpenLoops;
  # collider setup
  BEAM_1 2212; BEAM_ENERGY_1 6500;
  BEAM_2 2212; BEAM_ENERGY_2 6500;
  # scales and PDFs
  SCALE_VARIATIONS 0.25,0.25 0.25,1. 1.,0.25 1.,1. 1.,4. 4.,1. 4.,4.;
  SCALES STRICT_METS{MU_F2}{MU_R2}{MU_Q2};
  EXCLUSIVE_CLUSTER_MODE 1;
  CORE_SCALE QCD;
  PDF_LIBRARY LHAPDFSherpa;
  PDF_SET NNPDF30_nnlo_as_0118;
  USE_PDF_ALPHAS 1;
  # parton shower and matching
  NLO_CSS_PSMODE 1;
  PP_HPSMODE 0;
  CSS_SCALE_SCHEME 20;
  CSS_EVOLUTION_SCHEME 30;
  CSS_IS_AS_FAC 1;
  # fusing
  SHERPA_LDADD SherpaFusing;
  USERHOOK Fusing_Direct;
  FUSING_DIRECT_FACTOR 1;
  # model and flavor parameters
  EW_SCHEME 3; GF 1.1663787e-5;
  MASS[5] 4.75; MASSIVE[5] 1;
  MASS[6] 172.5; WIDTH[6] 0; STABLE[6] 0;
  MASS[24] 80.385; STABLE[24] 0;
}(run)
(processes){
  Process 93 93 -> 6 -6 5 -5
  Order (*,0)
  CKKW sqr(100);
  NLO_QCD_Mode MC@NLO
  Loop_Generator LOOPGEN
  ME_Generator Amegic
  RS_ME_Generator Comix
  PSI_ItMin 20000
  End process
}(processes)
\end{lstlisting}
\clearpage
Listing~\ref{runcard-fragmentation} shows the Sherpa runcard used for the fragmentation
component. We do not include settings related to the top decay handling, which differ
between the stable and the decayed top analysis and are irrelevant for the
fusing procedure studied here.
\begin{lstlisting}[label=runcard-fragmentation,captionpos=b,
  caption={\sherpa runcard for the fragmentation component of the fusing.},
  linerange={1-33,44-45,49-59}]
(run){
  MI_HANDLER None;
  ME_SIGNAL_GENERATOR Comix Amegic LOOPGEN;
  EVENT_GENERATION_MODE Weighted;
  LOOPGEN:=OpenLoops;
  OL_PARAMETERS=ew_scheme 2 ew_renorm_scheme 1
  # collider setup
  BEAM_1 2212; BEAM_ENERGY_1 6500;
  BEAM_2 2212; BEAM_ENERGY_2 6500;
  # scales and PDFs
  SCALE_VARIATIONS 0.25,0.25 0.25,1. 1.,0.25 1.,1. 1.,4. 4.,1. 4.,4.;
  SCALES STRICT_METS{MU_F2}{MU_R2}{MU_Q2};
  EXCLUSIVE_CLUSTER_MODE 1;
  CORE_SCALE QCD;
  PDF_LIBRARY LHAPDFSherpa;
  PDF_SET NNPDF30_nnlo_as_0118;
  USE_PDF_ALPHAS 1;
  # parton shower and matching
  NLO_CSS_PSMODE 1;
  PP_HPSMODE 0;
  METS_BBAR_MODE 5;
  CSS_SCALE_SCHEME 20;
  CSS_EVOLUTION_SCHEME 30;
  CSS_IS_AS_FAC 1;
  # fusing
  SHERPA_LDADD SherpaFusing;
  USERHOOK Fusing_Fragmentation;
  FUSING_FRAGMENTATION_STORE_AS_WEIGHT 1;
  # model and flavor parameters
  EW_SCHEME 3; GF 1.1663787e-5;
  MASS[5] 4.75;
  MASS[6] 172.5; WIDTH[6] 0; STABLE[6] 0;
  MASS[24] 80.385; STABLE[24] 0;
  # tags for process setup
  NJET:=4; LJET:=2; QCUT:=30.;
}(run)
(processes){
  Process : 93 93 ->  6 -6 93{NJET};
  NLO_QCD_Mode 3 {LJET}; CKKW sqr(QCUT/E_CMS);
  ME_Generator Amegic {LJET};
  RS_ME_Generator Comix {LJET};
  Loop_Generator LOOPGEN;
  Order (*,0);
  Enhance_Function VAR{pow(max(sqrt(H_T2)-PPerp(p[2])-PPerp(p[3]),(PPerp(p[2])+PPerp(p[3]))/2)/30.0,2)} {3,4,5,6}
  End process
}(processes)
\end{lstlisting}
\clearpage

\clearpage
\bibliography{main}

\end{document}